\documentclass[twocolumn]{article}
\usepackage[a4paper, margin=1.8cm]{geometry}

\usepackage[font=footnotesize,labelformat=simple]{subcaption}

\usepackage{graphicx} 

\usepackage{tikz,pgfplots}
\pgfplotsset{compat=1.18}
\usepackage{cite}
\usepackage{picinpar}
\usepackage{booktabs,amsmath,amsfonts,amssymb}
\usepackage{url}
\usepackage{flushend}
\usepackage[utf8]{inputenc}
\usepackage{colortbl}
\usepackage{soul}
\usepackage{multirow}
\usepackage{pifont}
\usepackage{color}
\usepackage{alltt}
\usepackage{cite}
\usepackage{hyperref}
\usepackage{enumerate}
\usepackage{siunitx}
\usepackage{pbox}
\usepackage{xcolor}
\usepackage{comment}
\usepackage{amsthm}
\usepackage{orcidlink}
\usepackage{cuted}    % For using the widetext environment
\usepackage{xfrac}

\newtheorem{result}{Result}
\newtheorem{theorem}{Theorem}
\newtheorem{remark}{Remark}
\newtheorem{assumption}{Assumption}
\newcommand{\real}{\mathbb{R}}
\newcommand{\rank}[1]{\text{rank}\left( #1 \right)}
\newcommand{\norm}[1]{\left\lVert #1 \right\rVert}
\newcommand{\abs}[1]{\left\lvert #1 \right\rvert}
\newcommand{\Hinf}{{\mathcal{H}_{\infty}}}
\newcommand{\estim}{\mathcal{E}}
\newcommand{\Vpk}{V_{\rm pk}}
\newcommand{\Vin}{V_{\rm in}}
\newcommand{\kff}{k_{\rm FF}}
\newcommand{\vsw}{v_{\rm SW}}
\newcommand{\dsw}{d}
\newcommand{\fsw}{f_{\rm sw}}
\newcommand{\controller}{\mathcal{K}}

\title{Robust Load Disturbance Rejection in PWM DC-DC Buck Converters}

\author{%Simone Pirrera \and  Francesco Gabriele \and     Davide Lena \and     Fabio Pareschi \and     Diego Regruto  \and   Gianluca Setti 
Simone Pirrera$^{\orcidlink{0009-0000-8301-9337}}$ \and     Francesco Gabriele$^{\orcidlink{0009-0006-9976-5741}}$ \and     Davide Lena \and     Fabio Pareschi$^{\orcidlink{0000-0001-8777-7135}}$  \and     Diego Regruto$^{\orcidlink{0000-0002-2144-2786}}$   \and   Gianluca Setti$^{\orcidlink{0000-0002-2496-1856}}$ \thanks{F. Gabriele and F. Pareschi are with the DET, Politecnico di Torino, 10129 Torino, Italy (e-mail: \{francesco.gabriele, fabio.pareschi\}@polito.it). 
    S. Pirrera and D. Regruto are with the DAUIN, Politecnico di Torino, 10129 Torino, Italy (e-mail: \{simone.pirrera, diego.regruto\}@polito.it). 
    D. Lena is with STMicroelectronics s.r.l. -- Torino, Italy. (email: davide.lena@st.com).
	G. Setti is with CEMSE, King Abdullah University of Science and Technology (KAUST), Saudi Arabia (e-mail: gianluca.setti@kaust.edu.sa).
	}
}
\date{}
 
\begin{document}

\maketitle

\begin{abstract}
This paper presents a novel approach to robust load disturbance rejection in DC-DC Buck converters. We propose a novel control scheme based on the design of two nested feedback loops. First, we design the controller in the outer loop using $H_\infty$ optimal control theory, and we show, by means of $\mu-$analysis, that such a controller provides robust stability in the presence of uncertainty affecting the physical parameters of the circuit. Then, we introduce an inner feedback loop to improve the system's response to output load disturbances. As far as the inner loop is considered, we propose a novel load estimation-compensation (LEC) scheme, and we discuss under what conditions the insertion of such an inner loop preserves the robust stability of the entire control system. The LEC scheme is compared with the other two linear structures based on well-established disturbance rejection methods. The advantages of LEC in terms of both complexity of implementation and obtained performances are discussed and demonstrated by means of numerical simulation. Finally, we present experimental results obtained through the implementation of the proposed control scheme on a prototype board to demonstrate that the proposed approach significantly enhances disturbance rejection performances with respect to the approach commonly used in DC-DC buck converters.
\end{abstract}

\section{Introduction}

The Pulse-Width Modulated (PWM) Buck converter is a DC-DC converter that steps down an unregulated input voltage, delivering a constant output voltage. The output voltage supplies external units that require accurate regulation to operate correctly. Applications of Buck converters include photovoltaic systems \cite{walker2004cascaded}, electric vehicles \cite{pavlovsky2013assessment}, and medical electronics \cite{park2023off}, among many others.  

Owing to its simplicity, Voltage-Mode Control (VMC) is one of the most popular control approaches for DC-DC converters. It is traditionally implemented in practice via a third-order (type-III) compensation network \cite{lee_ti}, whose design is based on an averaged Linear-Time Invariant (LTI) small-signal model of the converter \cite{erickson2007fundamentals}.  
From the converter perspective, the supplied system is a time-varying load current that may unpredictably change in time, causing output voltage fluctuations. Additionally, input supply (line) voltage may change, further degrading the output voltage tracking accuracy. Therefore, load and line variations are disturbances, and attenuating their effect on the output is a crucial aspect of the DC-DC converter design. In traditional VMC design, the Disturbance Rejection (DR) requirement contrasts with minimizing the propagation of switching noise caused by PWM \cite{erickson2007fundamentals}. Therefore, the controller must provide a trade-off that may result in an unsatisfactory response to disturbances.

Several techniques are available to improve DR performance. A popular technique in this context is the Feed-Forwarding (FF) \cite{Kazimierczuk_1996, Kazimierczuk_2000, Suntio_2007, Amer_Access_2024}, which is based on feedforward cancellation of the disturbance, which is assumed to be physically measured. Since the input voltage can be easily sensed, line-disturbance FF is feasible and implemented in commercial products; see, e.g., \cite{ti_tps40345, st2024l3751}. In contrast, implementing load-disturbance FF is impossible because the output current cannot be directly measured.

Within digital control, effective DR methods are based on adaptive schemes; see, e.g. \cite{Vongkoon_IEECON_2024, Bonizzoni_TCASII_2020, Yang_ACCESS_2023, Kim_ACCESS_2024}. Moreover, hybrid and time-varying model predictive control have been proven effective for DR in \cite{marie10_mpc} and \cite{liu18_mpcfpga}, respectively. However, these approaches incur significant additional costs and require increased circuit area due to the inclusion of digital components such as analog-to-digital converters and sample-and-hold circuits \cite{erickson2007fundamentals}. For this reason, analog solutions are preferable in several application scenarios. Among them, we mention \cite{Liu_2019_IFEEC, Zhou_TCASI_2020, Zeng_ISCAS_2019, Roh_tcasII}, which proposes alternative transistor-level architectures of Operational AMPlifiers (OpAMPs) and operational transimpedance amplifiers used to implement the traditional controller \cite{lee_ti}. Other analog-based DR methods include \cite{He_ISCAS_2022, Hsu_TCASII_2019, Yang_2020, Lee_2012}, which introduce additional components to modify either the feedback signal or the PWM pulse depth in response to load variations. However, the effects of the proposed modifications on the feedback system's theoretical properties are not analyzed. Their validity is solely confirmed through empirical simulations and experimental measurements. Furthermore, these approaches are architecture-specific, thus limiting their generalizability.

From a broader perspective, not limited to Buck converters, most DR techniques supported by rigorous theoretical analysis involve estimating the disturbance to apply a proper correcting action; see \cite{chen16tie} for a comprehensive review. Works \cite{wang15,yin17ies,wang2023intjcontrol} estimate the disturbance utilizing an Extended State Observer (ESO) \cite{gao2001cdc}. To enhance performance, a generalization of the ESO called Generalized Proportional-Integral Observer (GPIO) is considered in \cite{wang_robust_2017,wang_gpi_2019,xiong_robust_2022}. The main drawback of GPIO is that its analog implementation is expensive because it consists of a high-order filter for which a potentially large number of OpAMPs is required, thus increasing the overall system complexity and cost. DR schemes for which a simple implementation is possible are investigated in  \cite{yang18,lu_reduced-order_2018}, where reduced-order observer design is proposed, and in \cite{korompili_linear_2025}, which considers Kalman filter-based design. %Nevertheless, these works do not provide a \revthree{circuital implementation} of the proposed DR method.}

Another crucial aspect to be considered in the design of Buck converters is the presence of uncertainty in the values of the electrical components, which may strongly affect the closed-loop system's stability. In fact, ensuring closed-loop stability for all allowed component values (i.e., robust stability) is crucial for the safe operation of the circuit. This problem has been considered for various DC-DC converters in, e.g., \cite{garcera2000novel,zhang2014robust}. 
However, the problem of ensuring the robust stability of Buck converters that include DR schemes has received little attention so far. The problem is marginally considered in \cite[Sec.~II-E]{lu_reduced-order_2018}, where the authors study the location of the poles by varying the capacitance value only. They disregard the presence of uncertainty on other Buck converter component values, such as inductance, winding resistance, and capacitor series resistance. Instead, \cite[Sec.~III-A]{korompili_linear_2025} evaluates the closed-loop gain and phase stability margins. Yet, it is widely established that these indices are unreliable indicators of robustness \cite{lurie_classical_2000}. In this work, we are interested in developing novel DR schemes with guaranteed robust stability against uncertainty in all the component values using established results on $\Hinf$ and $\mu$-analysis \cite{skogestad2005multivariable,zhou1998essentials}.

The main contributions of this paper are summarized here. 
\begin{itemize}
    % 1 -- H infinity
    \item Starting from a suitable formalization of performance requirements, we design an $\Hinf$ optimal controller. Unlike previous works on $\Hinf$ Buck control \cite{xian11,mandal18,patra2021h}, we show that the $\Hinf$ controller transfer function resembles that of the traditional controller commonly used in commercial Buck converters \cite{lee_ti}. Furthermore, we prove its robust stability through $\mu$-analysis \cite{skogestad2005multivariable}. 
    % 2 -- Proposed compensations
    \item We propose three different load DR architectures. First, we apply the Disturbance Observer (DOB) \cite{ohishi87_dob,guo2005intjcontrol} and Unknown Input Observer (UIO) \cite{johnson68tac,johnson1971tac} designs. Next, we introduce a third, original approach named Load Estimator-Compensator (LEC), tailored for Buck converters. While our primary focus is on VMC, the proposed methodology is versatile and can be applied to other Buck control methods, such as peak/valley voltage or current mode control \cite{erickson2007fundamentals}.
    \item We demonstrate that direct insertion of LEC into the system preserves nominal closed-loop stability. This crucial property is not guaranteed when DR is performed using DOB and UIO. Furthermore, we derive a sufficient condition for robust stability in the presence of the LEC, and we show that the LEC can be designed to fulfill such a condition.
    % 3 -- Analog implementation
    \item We present a low-complexity analog implementation of the proposed LEC, thus avoiding the overhead costs involved in DR schemes based on using digital solutions and GPIO.
    % 4 -- Simulation and Measurements
    \item We compare the performance of the three considered DR schemes by means of numerical simulation.
    \item We show the effectiveness of the proposed control scheme compared to the traditional approach commonly used in commercial Buck converters by performing experimental laboratory tests on a prototype board implementing the proposed solution.
\end{itemize}

The paper is organized as follows. 
% Part 1: model and Hinf
In Sec.~\ref{sec:buck_model}, we present the Buck converter and its mathematical model. 
In Sec.~\ref{sec:maincontrol}, we review the traditional VMC design for Buck converters and basic notions of $\Hinf$ optimal control. Next, we formulate the control objectives and discuss the $\Hinf$ design for the Buck controller, showing that it is closely related to the traditional voltage-mode controller.
% Part 2: Compensation
In Sec.~\ref{sec:lti}, we introduce and analyze the three proposed load DR solutions. Sec.~\ref{sec:lec_analysis} focuses on the proposed LEC, providing nominal and robust stability results.
% Part 3: implementations and results
Sec.~\ref{sec:experim_res} presents results from simulations comparing the three proposed architectures, describes the analog circuit implementation of the system using the LEC, and provides results from experimental measurements conducted on a prototype board. Finally, Sec.~\ref{sec:concl} draws conclusions.

\section{Buck Converter Model}
\label{sec:buck_model}

\begin{figure*}
    \centering
    \includegraphics[width=0.6\linewidth]{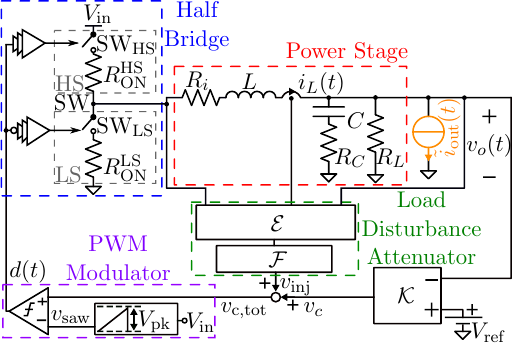}
    \caption{VMC Buck converter with load disturbance attenuator.}
    \label{fig:circuit}
\end{figure*}

\begin{figure}
    \centering
    \includegraphics[width=0.9\linewidth]{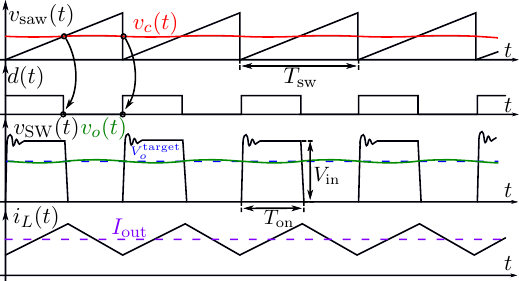}
    \caption{Steady-state operation of a Buck converter in CCM.}
    \label{fig:waveforms}
\end{figure}

\subsection{Architectural Description}

In this paper, we consider the Buck converter in Fig.~\ref*{fig:circuit}. From a high-level perspective, it consists of a feedback loop comprising the sawtooth-based PWM, the Half-Bridge (HB), the Power Stage (PS), and the controller $\controller$. At this stage, we neglect the presence of the inner loop implementing the load DR scheme, i.e., we neglect the presence of the two blocks $\estim$ and $\mathcal{F}$. We focus here on the outer control loop where the controller $\controller$ has to be designed to regulate the output voltage $v_o(t)$ to the target level $V_o^{\rm target}$.

The PWM stage implements a trailing-edge modulation. It compares the control signal $v_{\rm c, tot}(t)$ against $v_\textrm{saw}(t)$, which is a sawtooth signal with amplitude $V_\textrm{pk}$ and a fixed switching frequency of $f_{\rm sw} = \omega_{\rm sw}/(2 \pi)= 1/ T_{\rm sw}$. Accordingly, the PWM output is a logic signal $d(t) \in \{0,1\}$ having the same frequency as $v_\textrm{saw}(t)$.
The HB includes the High Side (HS) and Low Side (LS) power MOSFETs. The HS pulls the SW node to the input voltage $V_{\rm in}$; conversely, the LS ties the SW node to the voltage reference. For our purposes, the HS and LS MOSFETs are modeled with the switches $\rm SW^{HS,LS}$ in series with their ON resistances $R_{\rm ON}^{\rm HS, LS}$. The signal $d(t)$ drives the switches through gate drivers. The supply voltage, $\Vin$, remains constant during the operation of the Buck, i.e., $\Vin(t) = \Vin$ for all $t \geq 0$, but its actual value is only known to lie in $[V_o^{\rm target}, V_{\rm in,max}]$. Concerning the input voltage, the PWM block implements FF compensation.
The PS includes the output filter, composed of the inductor $L$ and the output capacitor $C$, together with the respective parasitic resistances $R_{i}$ and $R_{c}$. Furthermore, the load-resistance $R_L$ fixes the DC output current $I_{\rm out}$.
Finally, the controller $\mathcal{K}$ compares the constant voltage reference $V_\textrm{ref}$ with $v_o(t)$ to achieve the desired regulation by generating the command input $v_c(t)$. A detailed discussion about $\mathcal{K}$ is postponed to Sec.~\ref{sec:maincontrol}.

In this work, we will consider the Continuous Conduction Mode (CCM) operation of the Buck converter. In other words, we assume that during each period $[kT_\textrm{sw},(k+1)T_\textrm{sw}),\, k \in \mathbb{N}$, the inductor current is always positive; see, e.g., \cite{erickson2007fundamentals}. The Buck operating principle in CCM is sketched in Fig.~\ref*{fig:waveforms}. At steady-state, the circuit time evolution is cyclostationary and the duty-cycle of $d(t)$ and $v_{\rm SW}(t)$ is a constant value $D_c$. The value $D_c$ is such that the mean of the output voltage $v_o(t)$ equals $V_o^{\rm target}$. Conversely, when an exogenous disturbance is applied (e.g., a load current ${i}_o(t)$), $v_o(t)$ deviates from $V_o^{\rm target}$ and, consequently, $d(t)$ changes through the control action $v_c(t)$.

Throughout the paper, we will consider a case study to support the introduced arguments with numerical values. The considered parameter values and their uncertainties are listed in Table~\ref{tab:case_study}. The nominal load resistance and its uncertainty are defined respectively as $R_{L,\textrm{nom}} = (\underline{R_L}+\overline{R_L})/2$ and $\mathcal{I}_{R_L} \doteq [\underline{R_L},\overline{R_L}]$, where
\begin{equation}
    \underline{R_L} = \frac{V_o^{\rm target}}{I_\textrm{max}}, \quad \overline{R_L} = \frac{2L_\textrm{min} f_\textrm{sw}}{1-V_o^{\rm target}/V_\textrm{in,max}},
\end{equation}
and $L_{\rm min}$ is the minimum inductor value considering its uncertainty. The interval $\mathcal{I}_{R_L}$ guarantees the converter operation in CCM \cite{erickson2007fundamentals}.

\begin{table*}
\caption{Main Buck converter parameter's nominal values (Nom. Val.) and their respective uncertainties (Unc.), when applicable.}
\centering
    \begin{tabular}{c c c || c c c || c c c}
        \hline
        {Parameter} & {Nom. Val.} & {Unc.} & {Parameter} & {Nom. Val.} & {Unc.} & {Parameter} & {Nom. Val.} & {Unc.} \\ \hline           
        $C$&$\SI{0.249}{\milli\farad}$&$10\%$ & $R_C$&$\SI{0.115}{\milli\ohm}$&$15\%$ & $R_{\rm ON}^{\rm HS, LS}$& $\SI{6.5}{\milli\ohm}$ &$\pm 15\%$ \\ \hline
        $L$&$\SI{8.2}{\micro\farad}$&$20\%$ & $R_i$&$\SI{7}{\milli\ohm}$&$\pm 15\%$ & $I_\textrm{max}$&$ \SI{10}{\ampere}$&-\\ \hline
        $R_L$& $R_{L,\textrm{nom}}$&$\mathcal{I}_{R_L}$ & $\fsw$&$\SI{500}{\kilo\hertz}$& - & $V_\textrm{in,max}$&$\SI{20}{\volt}$&-\\ \hline
        $\kff$ &$30$&- & $V_{o}^{\rm target}$ & $\SI{5}{\volt}$ & - & & & \\ \hline
    \end{tabular}      
    \label{tab:case_study}
\end{table*}

\subsection{Mathematical model}
We model the Buck converter as the cascade connection of two blocks: the first one comprises the PWM and HB stages, while the second is the PS. This section presents the mathematical models of these blocks.

\subsubsection{PWM stage and Half Bridge} \label{sec:pwm_modul} 

The PWM stage is a nonlinear time-varying system implementing the function
\begin{equation}
    \dsw(t) = \textrm{PWM}(v_c(t)) = \begin{cases}
        0  \quad & v_c(t) < v_\textrm{saw}(t) \\ 
        1  \quad & \text{otherwise}.
    \end{cases}
\end{equation}

Motivated by the frequency-domain nature of the $\Hinf$ design, we study the frequency behavior of the PWM block. Results in this matter are available in the literature. Our analysis is based on the following result from \cite{sun2012pulse}. 

\begin{result}[Trailing-edge modulation of a sinusoid]
    Let's define the signal $v_1(t) = R_0+R_1 \cos(\omega_1 t + \theta_1)$. Suppose that $\omega_1 < \omega_{\rm sw}$ and that $R_0+R_1 < \Vpk$, $R_0-R_1 > 0$. The corresponding modulated signal $\dsw(t)=\textrm{PWM}(v_1(t))$ is given by: 
    
    \begin{subequations} \label{eq:vsw}
    \begin{align}
        & \dsw(t) = D_c+\frac{M}{2} \cos \left(\omega_1 t+\theta_1\right) + \label{eq:vsw_part1} \\
        &~~  + \sum_{m=1}^{+\infty} \frac{1}{m \pi} \left\{ \sin \left[m\left(\omega_\textrm{sw} t+\theta_c\right)\right] \right. + \label{eq:vsw_part2}  \\
        & \quad \left. - J_0(m \pi M) \sin \left[m\left(\omega_\textrm{sw} t+\theta_c\right)-2 m D \pi\right] \right\} + \notag \\
        &~~ + \sum_{m=1}^{+\infty} \sum_{n= \pm 1}^{ \pm \infty} \frac{J_n(m \pi M)}{m \pi} \times \label{eq:vsw_part3} \\
        & \quad \times \sin \left[\frac{n \pi}{2}-m\left(\omega_\textrm{sw} t+\theta_c\right)-n\left(\omega_1 t+\theta_1\right)+2 m D \pi\right] \notag
    \end{align} \end{subequations}
    where $J_n(z)$ are Bessel functions of the first kind, $D_c = R_0/V_{\rm pk}$ is the average duty ratio, and $M = 2R_1/\Vpk$ is the modulation index. 
\end{result}

Let us now analyze each term of Eq.~\eqref{eq:vsw} separately. 

\begin{enumerate}
    \item $D_c + \frac{M}{2} \cos \left(\omega_1 t+\theta_1\right)$ in Eq.~\eqref{eq:vsw_part1} is equivalent to $(1/{\Vpk}) {v}_1(t)$. 
    
    \item the terms in Eq.~\eqref{eq:vsw_part2} are frequency components at multiples of $\omega_\textrm{sw}$. Since $J_0(z) < 1$ for all $z \in \real$ \cite{watson1922treatise}, each term of this series is bounded in magnitude by $D_{m,0} \doteq 2/({m\pi})$.
    
    \item the terms in Eq.~\eqref{eq:vsw_part3} amount for the frequency components around $m\, \omega_\textrm{sw}$. Their amplitudes depend on the modulation index $M$ and, thus, on the amplitude $R_1$. However, using the property $\vert J_n(z) \vert \leq z^n/({n!}\,{2}^n)$ \cite[Page 49]{watson1922treatise} and given an upper bound $\overline{R}_1$ on $R_1$, we can bound the amplitudes of these terms as:
    \begin{equation}
        \frac{1}{m \pi n!}\left(\frac{m \pi M}{2}\right)^n \leq \frac{(m \pi)^{n-1}}{n!}\left(\frac{\overline{R}}{\Vpk}\right)^n \doteq D_{m,n}.
    \end{equation}
\end{enumerate}

Overall, an alternative representation of \eqref{eq:vsw} is 
\begin{subequations} \label{eq:dsw_lin_tot} \begin{align}
    \dsw(t) &= \frac{1}{\Vpk} {v}_1(t) + e_1(t) \label{eq:dsw_lin} \\
     e_1(t) &\doteq \sum_{m=1}^{+\infty} \sum_{n= -\infty}^{+\infty} A_{m,n} \cos[ \omega_{m,n} t + \phi_{m,n}]\label{eq:dsw_err}
\end{align} \end{subequations}
with $\omega_{m,n} \doteq m \omega_\textrm{sw} + n \omega_1$, $\phi_{m,n} \in [0,2\pi)$, and $A_{m,n} \leq D_{m,n}$. The bounds on $A_{m,n}$ are independent of $v_1$. Thus, as far as the controller design is concerned, we can consider each term of Eq.~\eqref{eq:dsw_err} as a sinusoidal disturbance at frequency $\omega_{m,n}$ that needs to be rejected.

Based on Eq.~\eqref{eq:dsw_lin_tot}, the PWM block allows for a linear model when a sinusoid is applied as input. Since the command input $v_c(t)$ does not contain a single harmonic, we look for a more general description. Consider the following assumptions:

\begin{assumption}\label{ass:vc_sum}
    The control input voltage, $v_c(t)$, is described by:
    \begin{equation}\label{eq:ass_vcseries}
        v_c(t) = R_{0} + \sum_{k \in \mathcal{I}_c} R_k \cos(\omega_k t + \theta_k) 
    \end{equation}
    where $\mathcal{I}_c$ is a (possibly infinite) set of indices such that $\omega_k < \omega_{\rm sw}$ and $\theta_k \in [0,2\pi)$.
\end{assumption}

\begin{assumption}\label{ass:smallsignal}
The PWM operates in a small-signal regime, i.e., the variations around the operating point $(R_0, D_c)$ are small. Therefore, there exists a constant $\overline{R}$ such that $R_k \leq \overline{R}$ for all $k$.
\end{assumption}

\begin{remark}\label{rem:crossover_freq}
    Assumption \ref{ass:vc_sum} is reasonable as long as the closed-loop system's bandwidth is lower than $f_\textrm{sw}$. This, therefore, leads to an easy-to-check design constraint.
\end{remark}

According to Assumption~\ref{ass:smallsignal} and the linear description \eqref{eq:dsw_lin_tot}, we can apply superposition of effects. 
We define $e_k$ according to \eqref{eq:dsw_err} with $\omega_k$ in place of $\omega_1$, and allowing $A_{m,n}$, $\theta_{m,n}$ to depend on $k$. Accordingly, we have 
\begin{equation}\label{eq:dsw_final}
    \dsw(t) = \frac{1}{\Vpk} v_c(t) + \sum_{k \in \mathcal{I}_c} e_{k}.
\end{equation}

\begin{remark}\label{rem:trad_model}
    The model in Eq.~\eqref{eq:dsw_final} is coherent with the averaged small-signal model of the PWM stage. Indeed, the latter coincides with the first term in Eq.~\eqref{eq:dsw_final}, i.e., $1/\Vpk$; see \cite{erickson2007fundamentals}. 
\end{remark}

\begin{remark}\label{rem:bndRk}
    Assumption \ref{ass:smallsignal} is compatible with the design constraint imposing that $v_c(t)$ never exceeds the voltage saturation limits imposed by the PWM stage, i.e., 
    \begin{equation}
        0 \leq v_c(t) \leq V_\textrm{pk}, \quad \forall t \geq 0.
    \end{equation}
    In fact, to meet this requirement, the sum of all contributions in~\eqref{eq:ass_vcseries} for $k \in \mathcal{I}_c$ must remain bounded by $\min(R_0,\Vpk-R_0)$.
\end{remark}

Next, the logic signal $\dsw(t)$ drives the HB block, which generates $\vsw(t) = \Vin \dsw(t)$. Considering Eq.~\eqref{eq:dsw_final}, we have
\begin{equation}\label{eq:vsw_model}
    \vsw(t) = \Vin \frac{1}{\Vpk} v_{c}(t) + d_a(t) 
\end{equation}
where $d_a(t)=\sum_{k \in \mathcal{I}_c} \Vin e_k(t)$ represents the total actuator disturbance to be rejected.
The magnitude of each harmonic of $d_a(t)$, i.e., $\Vin A_{m,n,k}$ is bounded by $V_{\rm in} D_{m,n}$, for all $m,n$ and $k$. Eq.~\eqref{eq:vsw_model} depends on the converter supply voltage $V_{\rm in}$, which is highly uncertain in practice. Since the considered Buck implements input voltage FF, this dependency is actually eliminated. In practice, the PWM stage is designed such that $V_{\rm pk}$ is adjusted according to $V_{\rm pk} = V_{\rm in}/\kff$, where $\kff$ is a constant called FF static gain.
Therefore, Eq.~\eqref{eq:vsw_model} simplifies to 
\begin{equation}\label{eq:vsw_final}
    \vsw(t) = \kff v_{c}(t) + d_a(t).
\end{equation}
Eq.~\eqref{eq:vsw_final} is a linear model of the PWM and HB blocks. %Additionally, this model is independent of $V_{\rm in}$, thus allowing for an improved line disturbance rejection.

\subsubsection{Power stage} \label{sec:power_stage} The PS is a multivariable LTI system described by:
\begin{equation}\label{eq:plant_PS}
    \begin{bmatrix} v_o(s) \\ i_L(s) \end{bmatrix} = 
    P(s) \begin{bmatrix} v_{\rm SW}(s) \\ \tilde{i}_{\rm out}(s) \end{bmatrix}=
    \begin{bmatrix} P_{11}(s) & P_{12}(s) \\ P_{21}(s) & P_{22}(s) \end{bmatrix} \begin{bmatrix} v_{\rm SW}(s) \\ \tilde i_\textrm{out}(s) \end{bmatrix}
\end{equation} 
where $v_{\rm SW}(t)$ is given in Eq.~\eqref{eq:vsw_final}, $\tilde i_\textrm{out}(t)$ is the output current disturbance, and $i_L(t)$ and $v_o(t)$ are the measured inductor current and output voltage, respectively. Applying standard circuit theory, we obtain:
\begin{subequations}\label{eq:plant_tfs} \begin{align}
    P_{11}(s) &= \frac{R_L(1+C R_c s)}{\alpha_0 s^2 + \alpha_1 s + \alpha_2} = P_{22}(s) \label{p11_22} \\
    P_{12}(s) &= R_L \frac{- C L R_c s^2 - (L + C R_c R_{i}') s - R_i'}{\alpha_0 s^2 + \alpha_1 s + \alpha_2}\label{p12}\\
    P_{21}(s) &= \frac{C (R_L + Rc) s + 1}{\alpha_0 s^2 + \alpha_1 s + \alpha_2}\label{p21}
\end{align}\end{subequations}
with 
\begin{equation}\begin{aligned}
    & R_{i}' = R_i + R_{\rm ON}^{\rm HS, LS}, \quad \alpha_0 = C L (R_L + R_c), \\
    & \alpha_1 = L + C R_L (R_c + R_{i}') + C R_c R_{i}', \quad \alpha_2 = R_L + R_{i}'
\end{aligned} \end{equation}
In the following, we denote as $\omega_{\rm ESR} = 1/(C R_c)$ the frequency of the zero of $P_{11}$ and as $\omega_\textrm{PS} = \sqrt{\alpha_2/\alpha_0}, \zeta_\textrm{PS}=\alpha_1/(2\sqrt{\alpha_0 \alpha_2})$ the frequency and damping of the plant poles, respectively.

\section{Design of the Voltage-Mode Controller $\mathcal{K}$}
\label{sec:maincontrol}
In this section we review the traditional design of $\controller$ (\ref{sec:model}) and basic notions of $\Hinf$ optimal control (\ref{sec:revHinf}). Next, we formalize the control design objectives and present an $\Hinf$ design of $\controller$ (\ref{sec:Hinf_design}), demonstrating its close connection with the VMC traditional design approach.

\subsection{Traditional Design Approach}
\label{sec:model}

\begin{figure}
    \centering
    \includegraphics{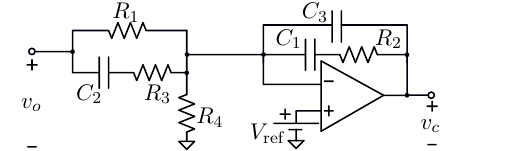}
    \caption{Type-III compensation network implementing $\mathcal{K}$.}
    \label{fig:type_III}
\end{figure}

The traditional design of $\controller$ is based on the availability of an averaged small-signal model of the Buck converter, i.e., an LTI model of the non-linear circuit derived under small-signal assumption. The small-signal averaging technique is a well-consolidated approach to derive LTI models of the PWM and the HB blocks, which constitute the only non-linear part of a Buck converter; see, e.g., \cite{erickson2007fundamentals,Lee_2006}. Specifically, the overall small-signal model of the plant (i.e., HB, PWM, and PS) is the transfer function:
\begin{equation}
    G_{v_c}^{v_{\rm o}}(s) = \frac{v_o{(s)}}{v_c(s)} = \kff P_{11}(s).
    \label{eq:hvd}
\end{equation}
In the traditional VMC design approach, which is widely adopted in commercial Buck converter implementations, $\controller$ is realized by the type-III compensation network shown in Fig.~\ref{fig:type_III}. We can easily derive the transfer function of the traditional type-III controller using circuit theory: 
\begin{equation}
    \mathcal{K}_\textrm{trad}(s) = \frac{G_{\rm c0} \left(1+s/{\omega_{\rm z,0}}\right)\left(1+s/\omega_{\rm z,1}\right)}{s\left(1+s/\omega_{\rm p,0}\right)\left(1+s/{\omega_{\rm p,1}}\right)},
    \label{eq:gc}
\end{equation}
where $G_{c0} = {R_1^{-1} (C_2+C_3)^{-1}}$, $\omega_{p,0} = {(R_3C_2)^{-1}}$, $\omega_{p,1} = {R_2^{-1}\frac{C_1+C_3}{C_1C_3}}$, $\omega_{z,0} = {(R_2C_1)^{-1}}$, and $\omega_{z,1} = {C_2^{-1}(R_1+R_3)^{-1}}$. 
Several design methodologies have been established to achieve both stability and the desired performances in VMC Buck converters~\cite{VenableTHEKF, kapat_2020, Wu_2010, Huang_Access_2024, Amer_Access_2024}. Typically the zeros $\omega_{z,0},\omega_{z,1}$ are placed close to the frequency $\omega_\textrm{PS}$, while the poles $\omega_{p,0} = \omega_{\rm sw}/2$ and $\omega_{p,1} = \min\{\omega_{\rm ESR}, \omega_{\rm sw}/2 \}$ serve to both suppress the switching-frequency noise and cancel out the effect of the $\omega_{\rm ESR}$ zero in~\eqref{eq:hvd}. Finally, the pole in $s=0$ ensures zero steady-state output voltage regulation error. Furthermore, $G_{c,0}$ is designed to set the closed-loop system bandwidth well-below $\omega_{\rm sw}$ and above $\omega_\textrm{PS}$. 

\subsection{Review of $\Hinf$ optimal control and $\mu$-analysis}\label{sec:revHinf}
This section reviews basic notions from $\Hinf$ optimal control theory. We refer the reader to, e.g., %\cite[Sec. 9.3]{skogestad2005multivariable}, 
\cite{skogestad2005multivariable}, \cite{zhou1998essentials} for more details.

The $\Hinf$ norm of a system $G(s)$ is defined as 
\begin{equation}\label{eq:hinf_def}
    \norm{G(s)}_\Hinf \doteq \sup_{\omega \in \real} \sigma_x \left( G(i\omega) \right) 
\end{equation}
where $\sigma_x(A)$ denotes the maximum singular value of the matrix transfer function $A(s)$.
The $\Hinf$ norm is the system gain induced by the $\ell_2$ to $\ell_2$ norm for signals, i.e., if $z = G w$, then
\begin{equation}
    \norm{z}_2 \leq \norm{G}_\Hinf \norm{w}_2
\end{equation}
where $\norm{x(t)}_2 \doteq \sqrt{\int_0^\infty \norm{x(t)}_2^2 \textrm{d}t}$. Accordingly, the physical meaning of the $\Hinf$ norm is the maximum energy amplification of the system.

Consider the feedback control system depicted in Fig.~\ref{fig:genplant} with plant $G(s) = \kff P_{11}$, controller $K(s)$, and feedback gain $G_f$. The $S/T$ mixed-sensitivity $\Hinf$ optimal control problem is given by \cite{skogestad2005multivariable,zhou1998essentials}:
\begin{equation}\label{hinf_opt}
    \gamma^* = \min_{K} \left\lVert \begin{bmatrix}
        W_1 S \\ W_2 T
    \end{bmatrix} \right\rVert_{\Hinf}
\end{equation}
where, $S(s) = (1+\mathcal{K}(s)\,G(s)\,G_f)^{-1}, T(s) = 1-S(s)$. The transfer functions $W_1(s), W_2(s)$ account for performance specifications by defining upper bounds on the desired magnitude of $S$ and $T$, respectively. The model of the plant, when augmented with $W_1, W_2$ to generate fictitious outputs $z_1 = W_1 S r$ and $z_2 = W_2 T r$, is referred to as the generalized plant.

In this work, we devote special attention to robust stability, i.e., the property of the feedback system to be stable for all allowed perturbations of certain parameters. Uncertainty is usually described by means of the feedback interconnection of a known block $M$ and an uncertain but bounded block $\Delta$, called $M-\Delta$ structure (see, e.g., \cite{skogestad2005multivariable,zhou1998essentials} for details). If $\Delta$ is such that $\norm{\Delta}_\Hinf < 1$, we deal with unstructured uncertainty, and robust stability is ensured if 
\begin{equation}\label{eq:mdelta_stability}
    \sup_{\norm{\Delta}_\Hinf <1} \norm{M \Delta}_\Hinf = \norm{M}_\Hinf < 1.
\end{equation}
When $\Delta$ is a matrix with a specific structure, e.g., when many independent sources of uncertainty are present, we deal with structured uncertainty. In this case, robust stability is ensured by looking at the structured singular value defined as
\begin{equation}\label{eq:def_mu}
    \mu(M) \doteq (\min \left\{k_m \mid \operatorname{det}\left(I-k_m M \Delta\right)=0 \right\})^{-1}.
\end{equation}
If $\mu<1$, robust stability is ensured. Methods to compute bounds on $\mu$ are available; see, e.g., \cite{packard1993complex}.

\subsection{Control objectives and $\mathcal{H}_{\infty}$ design}
\label{sec:Hinf_design}

This section presents an $\Hinf$ optimal controller design for the Buck converter. We consider the following performance objectives: 
\begin{enumerate}[R1)]
    \item zero steady-state regulation error;\label{req1}
    \item rejection of the disturbance $d_a(t)$ introduced by the PWM stage as defined in Sec.~\ref{sec:pwm_modul}; \label{req2} 
    \item minimization of overshoot/undershoot and settling time in the presence of output current disturbance; \label{req3}
    \item robust stability against the uncertainty of the Buck component values.\label{req4}
\end{enumerate}

\begin{figure}
    \centering
    \includegraphics[width=0.9\linewidth]{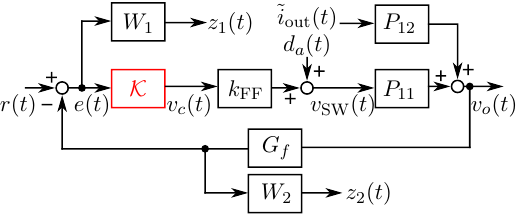}
    \caption{Generalized plant for $\Hinf$ design.}
    \label{fig:genplant}
\end{figure}

We formulate the problem in the $\mathcal{H}_\infty$ setting. Fig.~\ref{fig:genplant} shows the generalized plant used for design. 
We design the weighting functions $W_1(s),W_2(s)$ in Eq.~\eqref{hinf_opt} to account for the performance requirements R\ref{req1}-R\ref{req3}. Instead, we account for R\ref{req4} a-posteriori by proving robust stability of the designed controller through $\mu$-analysis. 

\begin{remark}
    In principle, we could account for R\ref{req4} \textit{a-priori} during the design phase in two ways. 
    The first is computing a bound on the unstructured uncertainty $\Delta$ and including the constraint $\norm{M}_\infty < 1$ in the $\Hinf$ optimization. Such a bound on $\Delta$ can be computed using the polynomial method in \cite{cerone2020bode}. However, this approach is conservative in presence of parametric uncertainty and leads to a closed-loop system with a very small bandwidth.
    The second approach is using $\mu$-synthesis \cite{packard1993complex} to account for the structured uncertainty. This approach is non-conservative but involves an iterative procedure (D-K iterations) that may converge to suboptimal solutions. Therefore, we prefer to assess the robust stability of the designed control loop \textit{a-posteriori} using $\mu-$ analysis.
\end{remark}

\begin{figure}
    \centering
        \input{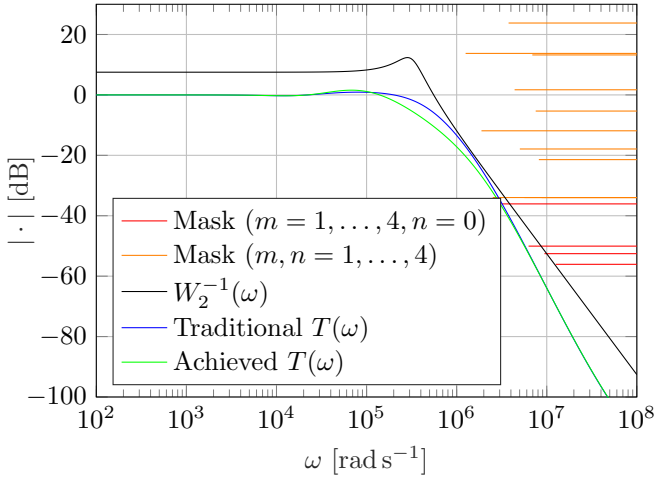}
    \caption{Magnitude Bode plot of $W_2^{-1}$ and complementary sensitivity achieved with traditional and $\Hinf$ controllers.
    Straight lines represent constraints \eqref{eq:cns_T_da}.}
    \label{fig:bodeT}
\end{figure}

We account for R\ref{req1}-R\ref{req3} by the following choices on $W_1, W_2$. To account for $R\ref{req1}$, we impose the presence of one pole at $s=0$ in $W_1$. For $R\ref{req2}$, given
\begin{equation}
    G_{d_a}^{v_c}(s) \doteq \frac{v_{c}(s)}{d_a(s)}= 
    %\frac{-P_{11} G_f \mathcal{K}}{1+\mathcal{K} \kff P_{11} G_f} = 
    \frac{1}{\kff} T(s),
\end{equation}
to reject the generic component of $d_a(t)$ at frequency $\omega = \omega_{m,n,k} = m \omega_{\rm sw} - n \omega_{k}$, we impose the constraint:
\begin{equation}\label{constrT}
    \abs{A_{m,n,k}} \abs{G_{d_a}^{v_c}(i\,\omega_k)} = \frac{D_{m,n} \Vin}{\kff} \abs{T(i\,\omega)} \leq \epsilon_{m,n} \Vin
\end{equation}
where $\epsilon_{m,n}$ is maximum error of $v_c(t)$ relative to $\Vpk$, for all $k$. Using $\Vpk = \Vin/\kff$, we get $\abs{T(i\,\omega)} \leq {\epsilon_{m,n}}/{D_{m,n}}$. The frequency components $\omega_{m,n,k}$ of $v_c(t)$ are mainly determined by the closed-loop crossover frequency $\omega_c$. As discussed in Remark~\ref{rem:crossover_freq}, we have that $\omega_k \leq \omega_c < \omega_\textrm{sw}/2$. At the same time, $\abs{T(i\,\omega)}$ is decreasing for $\omega \geq \omega_\textrm{sw}/2$. Consequently, $\omega_{\rm sw}/2$ is the worst-case over $k$ for constraint \eqref{constrT} and a sufficient condition for \eqref{constrT} is
\begin{equation}\label{eq:cns_T_da}
    \abs{T\left(i\,(m-n/2) \omega_\textrm{sw} \right)} \leq {\epsilon_{m,n}}/{D_{m,n}}.
\end{equation}
To ensure decaying error for increasing $m,n$, we select: $\epsilon_{1,0}=10^{-2}$, $\epsilon_{2,0}=10^{-3}$,  $\epsilon_{m+1,0}=0.5 \epsilon_{m,0}$ for $m \geq 2$, $\epsilon_{m,1} = 10^{-3}$ for all $m$, and $\epsilon_{m,n+1} = 0.5 \epsilon_{m,n}$ for all $m$ and $n \geq 1$. Eq.~\eqref{eq:cns_T_da} defines a set of constraints on the complementary sensitivity function $T$ (see, red and orange horizontal lines in Fig.~\ref{fig:bodeT}). 
    
Finally, to account for R\ref{req3}, we consider
\begin{equation}
    G_{\tilde i_\textrm{out}}^{v_o} \doteq \frac{v_o(s)}{\tilde i_\textrm{out}(s)} = %\frac{P_{12}}{1+K \kff P_{11} G_f} = 
    P_{12}(s) S(s).
\end{equation}
Thus, since $P_{12}(s)$ is fixed, R\ref{req3} is naturally recast as constraints on $S(s)$ through $W_1$. Specifically, the pole at $s=0$ in $W_1$ is sufficient to reject constant $\tilde i_\textrm{out}(t)$ at steady-state and a large bandwidth of $W_1$ is related to a fast response.

Overall, we select 
\begin{equation}\label{eq:weight_Hinf}
    W_1 = \frac{s^2+2 \zeta_\textrm{s} \omega_\textrm{s} s+\omega_\textrm{s}^2}{S_{p0} s (s+2 \zeta_\textrm{s} \omega_\textrm{s})}
    \quad
    W_2 = \frac{s^2+2 \zeta_t \omega_t s+\omega_t^2}{T_{p0} \omega_t^2}
\end{equation}
where $S_{p0} \approx T_{p0} \approx 2.5$, % Tp0 = 2.3765, Sp0 = 2.5703
$\zeta_\textrm{s} =\zeta_t = 0.3, \omega_\textrm{s} = 2 \omega_\textrm{sw},$ and $\omega_t =\omega_\textrm{sw}/10$.

\begin{remark}
    $W_1$ is the inverse of the sensitivity associated with a prototype second-order transfer function with natural frequency $\omega_\textrm{s}$, which is a high frequency for which $\tilde i_\textrm{out}(t)$ would be well-rejected.
\end{remark}

\begin{remark}
    $W_2$ is built such that its inverse satisfy constraints \eqref{eq:cns_T_da}. In other words, $\omega_t$ is such that, in Fig.~\ref{fig:bodeT}, the magnitude bode plot of $W_2^{-1}$ (\ref{plotline:W2inv}) is lower than the masks (\ref{plotline:maskcnsred},\ref{plotline:maskcnsorange}).
\end{remark}

 The optimization problem \eqref{hinf_opt} is solved using convex semidefinite programming optimization \cite{iwasaki1994all}. This way, we get a controller $K^{**}$ with $5$ poles $p_0,\dots,p_4$ and $5$ zeros $z_0,\dots,z_4$. Specifically: the poles are such that $p_0=0$, $p_1 \approx -\omega_\textrm{sw}$, $p_2 \approx -0.5 \omega_\textrm{sw}$, and $\abs{p_3},\abs{p_4} > \omega_\textrm{sw}$, while the zeros are such that $z_0,z_1$ exactly cancel the plant poles at $\omega_\textrm{PS}$ and $\abs{z_2},\abs{z_3},\abs{z_4}>\omega_\textrm{sw}$.
The optimal objective value achieved by this controller is $\gamma^* > 1$, which means that the requirements imposed by $W_1, W_2$ cannot be simultaneously fulfilled. This is mainly due to the choice of a large $\omega_s$. Indeed, R\ref{req2} asks for a low bandwidth of the closed-loop system, and, at the same time, R\ref{req3} requires a large bandwidth. This consideration strongly motivates inserting the disturbance rejection scheme discussed in Sec.~\ref{sec:lti}.

The controller $K^{**}$ is simplified by discarding high-frequency dynamics. This way, we obtain the third-order candidate controller
\begin{equation}\label{Kinf}
    K^*(s) = \frac{G (s^2 + 2 \zeta_\textrm{PS} \omega_\textrm{PS} s + \omega_\textrm{PS}^2)}{s (s+p_1) (s+p_2)}, \quad G \in \real.
\end{equation}

We notice that the zeros of \eqref{Kinf} are complex and conjugate; thus, their implementation using an analog circuit is critical. To overcome this issue, we replace them with real coincident zeros at $\omega_\textrm{PS}$, obtaining the final controller
\begin{equation}\label{Kfinal}
    \mathcal{K}(s) = \frac{G (s + \omega_\textrm{PS})^2}{s (s+p_1) (s+p_2)}.
\end{equation}
This is a valid approximation as long as the damping $\zeta_\textrm{PS}$ is not too small, which is true under in CCM.

\begin{figure}
    \centering
        \input{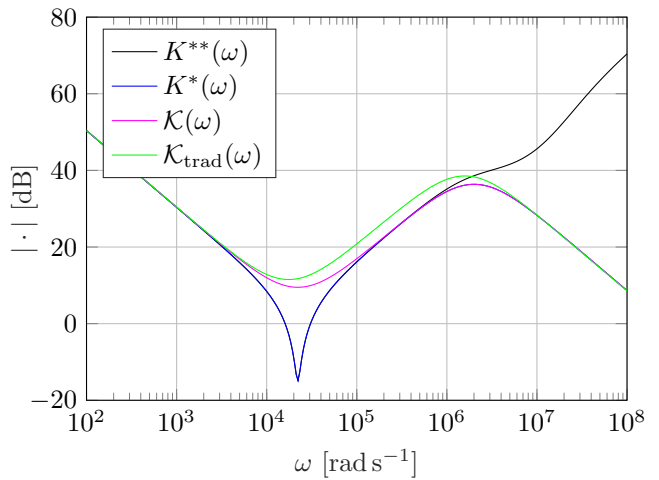}
    \caption{Controller magnitude Bode diagram}
    \label{fig:bodeGc}
\end{figure}

Fig.~\ref{fig:bodeGc} compares the magnitude Bode plot of the optimal order 5 controller, the simplified controller after the removal of high-frequency dynamics $K^*$, the final controller $\mathcal{K}$, and the traditional controller $\mathcal{K}_\textrm{trad}$. 

\begin{remark}
    Eq.~\eqref{Kfinal} and Fig.~\ref{fig:bodeGc} show that the structure of the final controller obtained through $\Hinf$ design coincides with that of the traditional VMC controller in \eqref{eq:gc}. Regarding the parameter values, the only difference is the frequency of the pole $\omega_{p,1}$, which is $\omega_{\rm sw}$ for the $\Hinf$ case and $\min\{\omega_{\rm ESR}, \omega_{\rm sw}/2 \}$ for the traditional VMC controller.
\end{remark}

\begin{figure}
    \centering
        \input{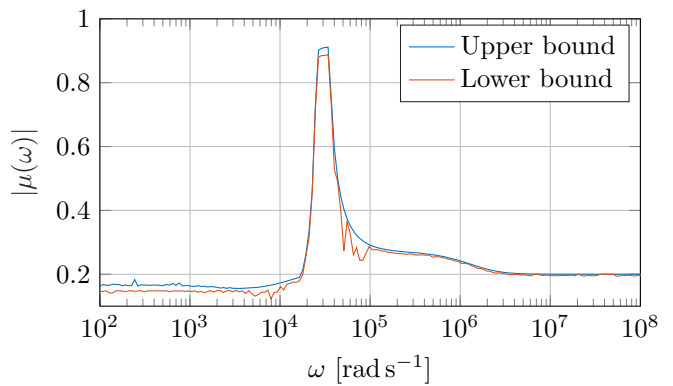}
    \caption{Bounds on $\mu$ for the final controller $\mathcal{K}$ in Eq.~\eqref{Kfinal}}
    \label{fig:mu}
\end{figure}

Finally, we account for the robust stability requirement R\ref{req4} by performing a $\mu$-analysis. We consider the controller \eqref{Kfinal} and the structured description of the uncertainty obtained from the component uncertainties. Fig.~\ref{fig:mu} shows the computed bounds on $\mu$ for the case study in Tab.~\ref{tab:case_study}. These results are obtained using MATLAB Robust Control Toolbox \cite{Balas_Chiang_Packard_Safonov_2015}. We observe that the most critical frequencies are around $\omega_\textrm{PS}$, where the frequency response of the plant is very sensitive to parameter variations. Since $\mu<1$ for all frequencies, we conclude that \eqref{Kfinal} achieves robust stability.

\section{Disturbance Rejection Techniques}
\label{sec:lti}
This section illustrates the design of three DR schemes. First, we present designs based on DOB and UIO methods. Next, we introduce an original approach called LEC, which combines the best features of the above two.

All three solutions in this section consist of the cascade of two blocks: an estimator $\mathcal{E}$ and a filter $\mathcal{F}$. The load disturbance is estimated via $\mathcal{E}$, while $\mathcal{F}$ compensates for its effects as if it were directly measured. The combination of $\mathcal{E}$ and $\mathcal{F}$ provides us with an additional feedback control action $v_{\rm inj}(t)$, which adds to $v_c(t)$ produced by $\mathcal{K}$. Fig.~\ref{fig:circuit} shows a diagram of the overall feedback control system.

\subsection{Disturbance observer design}
The design of the DOB (see, e.g., \cite{ohishi87_dob}) is based on the idea of describing the effect of disturbances and unmodeled dynamics as an unknown exogenous input disturbance $\delta(t)$ acting on the input in an additive way. Formally, we define an equivalent input signal $v_\textrm{SW,eq}(t)$ to account for the effect of the presence of the disturbance on the output, i.e., such that
\begin{align} \label{dob_dist_model}
    v_o(s) = P_{11}(s) v_\textrm{SW,eq}(s), \quad 
    v_\textrm{SW,eq}(t) \doteq \vsw(t) + \delta(t). 
\end{align}
The signal $\delta(t)$ has no physical meaning and accounts for all the mismatches between the model output and the measured one, independently of their source. 

Relying on measurements of the output $v_o(t)$ and the actual input $v_{\rm SW}(t)$, the disturbance is computed as
\begin{equation}\label{dob_dist_estim}
    \delta(s) = P_{11}^{-1}(s) v_o(s) - \vsw(s).
\end{equation}
However, since the transfer function $P_{11}(s)$ is strictly proper, high-frequency poles must be included to define a physically realizable approximation of $P_{11}^{-1}(s)$. In the DOB literature, this is addressed by introducing a filter $Q(s)$ having as many poles as the plant relative degree to estimate the disturbance by
\begin{equation}\label{dob_dist_estim2}
    \hat \delta(s) = Q(s) [ P_{11}^{-1}(s) v_o(s) - \vsw(s) ].
\end{equation}
In our case, $P_{11}(s)$ has relative degree one and the most natural choice is $Q(s) = {p_H}/{(s+p_H)}$, where $p_H$ is at high-frequency.

Furthermore, the implementation of $Q P_{11}^{-1}(s)$ with a simple analog circuit is challenging due to the presence of complex conjugate zeros: this is the same issue arising during the design of the $\Hinf$ optimal controller in Sec~\ref{sec:Hinf_design}. Therefore, an analog implementation of the DOB estimator requires approximating the complex conjugate zeros with real ones. Overall, we obtain:
\begin{align}\label{dob_final}
    \hat \delta(s) &= \estim(s) \begin{bmatrix}
        v_o(s) \\ \vsw(s)
    \end{bmatrix} = \begin{bmatrix}
        G_\textrm{DOB}(s) & -Q(s)
    \end{bmatrix} \begin{bmatrix}
        v_o(s) \\ \vsw(s)
    \end{bmatrix},
\end{align}
where
\begin{align}\label{eq:Gdob}
    G_\textrm{DOB}(s) \doteq \frac{(1+s/\omega_\textrm{PS})^2}{(1+s/\omega_{\rm ESR}) (1+s/p_H)}.
\end{align}
Next, we determine the compensating action $v_{\rm inj}(t)$ that cancels the effect of $\delta(t)$ on the output, i.e., such that
\begin{equation}
    \kff P_{11}(s) v_{\rm inj}(s) + P_{11}(s) \delta(s) = 0.  
\end{equation}
Therefore, we define $\mathcal{F}(s)$ as:
\begin{equation}\label{dob_control_law}
    v_{\rm inj}(s) = \mathcal{F}(s) \hat \delta(s) = - \frac{1}{\kff} \hat \delta(s).
\end{equation}
The DOB approach allows us to design the compensation mechanism using output voltage measurements only. 

We also notice that DOB does not rely on models describing how the disturbance affects the system output. On the one hand, this makes the approach appealing when it is hard, or even infeasible, to obtain a good model for the disturbance. On the other hand, when an accurate model of how the disturbance influences the output is available (as in the considered case), neglecting this information leads to over-compensation. That is to say, not only is the effect of the disturbance compensated, but also other potentially harmless effects, e.g., the effect of component uncertainty. In general, over-compensation is not problematic, but, as it will be shown in Sec.~\ref{sec:simul}, simulation examples demonstrate that in the presence of parameter uncertainty, the DOB compensation is associated with larger command input actions that potentially exceed the PWM stage saturation limits.
The over-compensation effect is additionally emphasized by the need to approximate $P^{-1}_{11}$ and to introduce $Q$. In fact, the estimate $\hat d$ will be non-zero even in the ideal case, i.e., when disturbances are absent and all components are perfectly known.

To complete our analysis of the DOB, we notice that with the considered DOB, the transfer function between $v_{c}(t)$ and $v_o(t)$ is not preserved. Accordingly, the theoretical properties ensured by $\mathcal{K}$ on the nominal plant, including robust stability, are lost. The robust stability of the feedback control systems incorporating DOB compensation is studied in \cite{sariyildiz2014guide}. Still, these results do not fully apply in our case due to the need to approximate $P_{11}^{-1}(s)$ to ease its analog implementation.

\subsection{Unknown input observer design}
\label{sec:uio}
The design of the UIO is based on the idea of using an observer to design the estimator $\mathcal{E}$. ESO and GPIO methods are based on the same idea but use different observers. 

The main underlying assumption is that the disturbance can be modelled as the output of an exo-system: 
\begin{equation}\label{exosys}
    \dot \delta(t) = A_d \delta(t), \qquad \tilde i_\textrm{out}(t) = C_d \delta(t)
\end{equation}
For the case of interest, we take the following assumption
\begin{assumption}\label{ass:io_const}
    The output current disturbance, $\tilde i_\textrm{out}(t)$, is almost constant, i.e., $\frac{\textrm{d}}{\textrm{d} t} \tilde i_\textrm{out}(t) = 0$ for almost every time $t$. 
\end{assumption}
Accordingly, $\tilde i_\textrm{out}(t)$ is the constant output of an exo-system with a pole in origin, i.e., $A_d=0$ and $C_d=1$ in Eq.~\eqref{exosys}.

Consider the state-space representation of the plant
\begin{equation}\label{plant_ss}
    \begin{bmatrix} v_o(s) \\ i_L(s) \end{bmatrix} = \left[\begin{array}{c|c c}
         A_p & B_{p,1} & B_{p,2} \\
         \hline
         C_{p} & D_{p,1} & 0_{2 \times 1}
    \end{array} \right] \begin{bmatrix} \tilde i_\textrm{out}(s) \\ v_{\rm SW}(s) \end{bmatrix}.
\end{equation}

Combining \eqref{plant_ss} with \eqref{exosys}, we define an augmented state-space description of the plant where the disturbance is regarded as an additional state:
\begin{equation}\label{plant_aug}
    \begin{bmatrix} v_o(s) \\ i_L(s) \end{bmatrix} = \left[\begin{array}{c|c}
         A_a & B_{a}  \\
         \hline 
         C_{a} & 0_{2 \times 1} 
    \end{array} \right] \begin{bmatrix} v_{\rm SW}(s) \end{bmatrix}
\end{equation}
where
\begin{equation}\label{aug_ss_mtx}\begin{aligned}
    A_a &= \begin{bmatrix} A_p & B_{p,1} \\ 0_{1 \times 2} & 0 \end{bmatrix} \in \real^{3\times 3}, \quad B_a = \begin{bmatrix} B_{p,2} \\ 0 \end{bmatrix} \in \real^{3 \times 1}, \\
    C_a &= \begin{bmatrix} C_{p} & D_{p,1} \end{bmatrix} \in \real^{2 \times 3}.
    \end{aligned}
\end{equation}

Using \eqref{plant_aug}, we design a Luenbergher observer $\mathcal{E}(s)$ to estimate the states of \eqref{plant_aug}. With this choice, $\mathcal{E}(s)$ is parametrized by a gain matrix $F \in \real^{3 \times 2}$ and described by
\begin{equation}\label{io_hat_sys}\begin{aligned}
        \hat{i}_\textrm{out}(s) & = \mathcal{E}(s) \begin{bmatrix} v_{\rm SW}(s) \\ v_o(s) \\ i_L(s) \end{bmatrix} = \\
        &= \left[ \begin{array}{c|c c}
         A_a - F C_a & B_a & F \\
         \hline
         C_o & 0 & 0_{1 \times 2} 
    \end{array} \right] \begin{bmatrix} v_{\rm SW}(s) \\ \begin{bmatrix} v_o(s) \\ i_L(s) \end{bmatrix} \end{bmatrix}
\end{aligned} \end{equation}
where $C_o = \begin{bmatrix} 0 & 0 & 1 \end{bmatrix}$ is used to select the estimate of the last state of \eqref{plant_aug} only, i.e., $\hat{i}_\textrm{out}(t)$. By design, $F$ is such that
\begin{equation}\label{eq:lamb_uio}
    (A_a - F C_a)U = U \Lambda
\end{equation}
for some invertible matrix $U \in \real^{3 \times 3}$ and with $\Lambda = \text{diag}([\lambda_1,\lambda_2,\lambda_3])$. If $\lambda_i < 0$ for all $i=1,2,3$, we have  
\begin{equation}
    \hat{i}_\textrm{out}(t) \rightarrow \tilde i_\textrm{out}(t), \quad \text{ as } t \rightarrow \infty.
\end{equation}
The convergence rate is determined by the choice of $\Lambda$. High frequency eigenvalues lead to a more accurate estimation of $\tilde i_\textrm{out}(t)$ and, consequently, better disturbance rejection performances. However, if the frequency is too high, the analog implementation of the filter is more challenging and eventually becomes infeasible.

According to the considerations in \cite[Section 2.2]{kautsky1985robust}, standard algorithms allow us to place at most $m$ coincident eigenvalues, where $m$ is the number of measured outputs. Therefore, in our case, thanks to the inductor current sensing, it is possible to choose $\lambda_2 = \lambda_3$. 
\begin{result}[Current estimator observability]\label{result1}
    The current estimator described by \eqref{io_hat_sys} has an unobservable mode. Consequently, $\mathcal{E}$ is exactly described by a row vector of $3$ transfer functions of dynamical order $2$ that share the same poles.
\end{result}
\begin{proof}
   We show that the rank of the observability matrix $O$ associated with the system $\mathcal{E}$ is $2$, i.e.,  
    \begin{equation}
        \rank{O} = \rank{\begin{bmatrix} C_o \\ C_o (A_a - F C_a) \\ C_o (A_a - F C_a)^2 \end{bmatrix}} = 2.
    \end{equation}
    Consider the eigendecomposition of $A_a-L C_a$: 
    \begin{equation}
        A_a - F C_a = \sum_{i=1}^3 \lambda_i u_i v_i^\top, \quad V = U^{-1}. 
    \end{equation}
    Let us now write the rows of the matrix $O$ as:
    \begin{equation}
    \label{row_eq}
    \begin{aligned}
        &C_o I = C_o U V = \sum_{i=1}^3 C_o u_i v_i^\top = \\
        &\quad = U_{1,3} v_1^\top + ( U_{2,3} v_2^\top + U_{3,3} v_3^\top )\\
        &C_o (A_a - F C_a) = \sum_{i=1}^3 \lambda_i C_o u_i v_i^\top = \\
        &\quad= \lambda_1 U_{1,3} v_1^\top + \lambda_2 ( U_{2,3} v_2^\top + U_{3,3} v_3^\top )\\
        &C_o (A_a - F C_a)^2 = \sum_{i=1}^3 \lambda_i^2 C_o u_i v_i^\top = \\
        &\quad= \lambda_1^2 U_{1,3} v_1^\top + \lambda_2^2 ( U_{2,3} v_2^\top + U_{3,3} v_3^\top )
    \end{aligned}
    \end{equation}
    Noting that the equations in \eqref{row_eq} are linearly dependent, the result follows. 
\end{proof}

According to the Result~\ref{result1}, it is possible to implement the estimator using second-order filters, which are cheaper to implement compared to third-order ones. Still, the UIO estimator requires three inputs, which makes it more expensive than the DOB.

\begin{remark}
    The estimator developed in this section is strongly related to the reduced-order enhanced state observer in \cite{lu_reduced-order_2018}. Indeed, an observer with the same structure is developed and reduced to the second order. Differently from \cite{lu_reduced-order_2018}, our observer exploits the sensing of $i_L(t)$ to enhance the estimation speed and accuracy. Moreover, we estimate $\tilde i_\textrm{out}(t)$ rather than a term accounting for all unmodeled disturbances, thus avoiding overcompensation effects. 
\end{remark}

Finally, the estimate $\hat{i}_\textrm{out}(t)$ is provided as input to a filter $\mathcal{F}(s)$, which must be designed such that the effects of $\tilde i_\textrm{out}(t)$ on the output are canceled. A detailed treatment on the design of $\mathcal{F}(s)$ is postponed to Sec.~\ref{sec:comp_design}.
Similarly to DOB, introducing the UIO does not preserve the transfer function between the control input and the output, thus losing the stability guarantee. 

\subsection{Load Estimator Compensator design}
\label{sec:lec}
In this section, we present an alternative disturbance estimator, named LEC, intended to mitigate the above-mentioned drawbacks of DOB and UIO. Specifically, unlike DOB, the LEC estimates the load current disturbance, thus avoiding overcompensation effects. Moreover, unlike UIO, the LEC allows for a cheap analog implementation (see Sec.~\ref{sec:circuital_implentation}). Finally, the LEC scheme allow us to derive robust stability conditions (see Sec.~\ref{sec:robustness}), which is a major advantage compared to both DOB and UIO.

The main idea is to estimate $\tilde i_\textrm{out}(t)$ through algebraic manipulations relying only on the knowledge of the measured outputs $v_o(t)$ and $i_L(t)$. In the following equations, we omit, for brevity, the Laplace argument $s$. First, we obtain $\vsw(s)$ as a function of $v_o(s)$ and $\tilde i_\textrm{out}(s)$ as
\begin{equation}
    v_{\rm SW} = \frac{1}{P_{11}} v_o - \frac{P_{12}}{P_{11}} \tilde i_\textrm{out}.
\end{equation}
Then, we replace into the Eq.~\eqref{eq:plant_PS} for $i_L(s)$:
\begin{equation}\label{iL_fcn_of_vo_and_io}\begin{aligned}
    i_L &= P_{21}v_{\rm SW} + P_{22}\tilde i_\textrm{out} = \\
    &= \frac{P_{21}}{P_{11}} v_o - \frac{P_{21} P_{12}}{P_{11}} \tilde i_\textrm{out} + P_{22}\tilde i_\textrm{out} = \\
    &= \frac{P_{21}}{P_{11}} v_o + \frac{P_{22}P_{11} - P_{21} P_{12}}{P_{11}} \tilde i_\textrm{out}.
\end{aligned}\end{equation}
Let us denote $\Delta_P \doteq P_{22}P_{11} - P_{21} P_{12}$. From \eqref{iL_fcn_of_vo_and_io}, we get
\begin{equation}\label{io_fcn_out}
    \tilde i_\textrm{out} = \frac{P_{11}}{\Delta_P}\left( -\frac{P_{21}}{P_{11}} v_o + i_L \right).
\end{equation}

Ideally, if the plant is exactly known, the right-hand side of Eq.~\eqref{io_fcn_out} can be used to define the filter $\mathcal{E}$, and $\hat i_\textrm{out} = \tilde i_\textrm{out}$ by construction. 

\begin{remark}
    Thanks to the algebraic design procedure, no assumptions on the nature of the current disturbance $\tilde i_\textrm{out}(t)$ are needed. This is particularly meaningful in comparison with observer-based solutions. Indeed, designs based on UIO and ESO are based on Assumption \ref{ass:io_const}, while GPIO design requires a suitable assumption on the $\tilde i_\textrm{out}$ derivatives; see, e.g., \cite{wang_robust_2017}.
\end{remark}

\begin{remark}
    Since the LEC design is based on algebraic cancellations, uncertainty on the plant parameters is critical, especially regarding the robust stability of the system. This point is extensively discussed in Sec.~\ref{sec:robustness}.
\end{remark} 

In the remainder of this section, we prove that the resulting estimator is described by a first-order proper transfer function. Thus, it can be physically implemented through a cheap analog circuit. Let us compute each term of \eqref{io_fcn_out} as a function of the Buck components.

\begin{result}[Characterization of $\Delta_P$]
    For a Buck converter, 
    \begin{equation}\label{deltaP}
        \Delta_P = P_{11}
    \end{equation}
    independently of the components' values. 
\end{result}
\begin{proof}
    We replace \eqref{p11_22}-\eqref{p21} into the definition of $\Delta_P$. Let us now evaluate the products $P_{11}P_{22}$ and $P_{12}P_{21}$ as a preliminary calculation. For ease of notation, we denote $D = \alpha_0 s^2 + \alpha_1 s + \alpha_2$. Since
    \begin{equation}\label{p12fact}
        P_{12} = -R_L\frac{(1+C R_c s)(sL+R_i')}{\alpha_0 s^2 + \alpha_1 s + \alpha_2},
    \end{equation}
    we have:
    \begin{equation}\begin{aligned}
        P_{11}P_{22} &= P_{11}^2 = \frac{R_L^2}{D^2} (1+C R_c s)^2
    \end{aligned}\end{equation}
    and
    \begin{equation}\begin{aligned}
        -P_{12}P_{21}& = \frac{R_L}{D^2}(1+C R_c s)(sL+R_i')\\
        &\qquad((C R_L + C Rc) s + 1).
    \end{aligned} \end{equation}
    Taking the sum of the above terms leads to 
    \begin{equation}\begin{aligned}
        \Delta_P &= \frac{R_L}{D^2} (1+C R_c s) (R_L + C R_L R_c s + \\
        &\quad +(sL+R_i')(C(R_L + Rc) s + 1)) = \\
        &= \frac{R_L}{D^2} (1+C R_c s) D = P_{11}
    \end{aligned} \end{equation}
\end{proof}

\begin{result}[Characterization of $P_{21}/P_{11}$]
For a Buck converter,
\begin{equation}\label{G1_nom}
    G_1 \doteq \frac{P_{21}}{P_{11}} = \frac{C(R_L+R_C)s+1}{R_L(1+CR_c s)}.
\end{equation}
\end{result}
\begin{proof}
    The result directly follows from Eqs.~\eqref{p11_22} and \eqref{p21}.
\end{proof}

Finally, replacing Eqs.~\eqref{deltaP} and \eqref{G1_nom} in the right-hand side of \eqref{io_fcn_out}, we define the output current estimator $\mathcal{E}$ as
\begin{equation}\label{eq:def_estim_lec}
    \hat{i}_\textrm{out} = \mathcal{E} \begin{bmatrix} v_o \\ i_L \end{bmatrix} = \begin{bmatrix} -G_1 & 1 \end{bmatrix} \begin{bmatrix} v_o \\ i_L \end{bmatrix}.
\end{equation}

\begin{remark}
    $\mathcal{E}$ is a proper transfer function and can be physically implemented using an analog circuit. See Sec.~\ref{sec:circuital_implentation}.
\end{remark}

\subsection{Design of compensation filter for UIO and LEC}
\label{sec:comp_design}
This section discusses how to design the compensation filter $\mathcal{F}$ assuming that the estimator stage $\estim$ provides a correct estimate $\hat i_\textrm{out}$ of the output current disturbance $\tilde i_\textrm{out}(t)$. Ideally, the filter $\mathcal{F}$ should inject an additional control input $v_{\rm inj}(t)$ such that the effect of $\tilde i_\textrm{out}(t)$ on $v_o(t)$ is null, i.e., 
\begin{equation}\label{comp_eq}
    \kff P_{11} v_{\rm inj} + P_{12} \tilde i_\textrm{out} = 0.
\end{equation}
From Eq.~\eqref{comp_eq}, a natural candidate for defining $\mathcal{F}$ would be $-{P_{12}}{\kff^{-1} P_{11}^{-1}}$.

\begin{result}[Characterization of $P_{12}/P_{11}$] \label{result:p12divp11}
For a Buck converter,
    \begin{equation}\label{G2_nom}
        G_2 \doteq \frac{P_{12}}{P_{11}} = -sL - R_i - R_{\rm ON}^{\rm HS, LS}.
    \end{equation}
\end{result}
\begin{proof}
    The result follows by replacing $P_{12}$ with \eqref{p12fact} and $P_{21}$ with \eqref{p21}, and performing suitable simplifications.
\end{proof}

Result \ref{result:p12divp11} allows us to conclude that $\mathcal{F}$ cannot be defined according to $-\kff^{-1}G_2$ because it is not a proper transfer function. To address this issue, we define $\mathcal{F}$ as
\begin{equation}\label{eq:def_compens}
    \mathcal{F} \doteq - \frac{G_2}{\kff} \frac{1}{1+s/p_H} = \frac{sL + R_i + R_{\rm ON}^{\rm HS, LS}}{\kff(1+s/p_H)}.
\end{equation}
where $p_H > 0$ is a high-frequency pole. The rationale of this choice is that as $p_H \rightarrow \infty$, the original desired frequency behavior is recovered. 

\begin{remark}
    As we will see in Sec.~\ref{sec:robustness}, the frequency of the pole $p_H$ must be tuned to ensure a suitable trade-off between output current DR performance and robustness.
\end{remark}

%%%%%%%%%%%%%%%%%%%%%%%%%%%%%%%%%%%%%%%%%%%%%%%%%%%%%%%%%%%%%%%%%%%%%%%%%%%%%%%%%%%%%%%%%%%%%%%%%%%%%%%%%%%%%%%%%%%%

\section{LEC stability analysis}
\label{sec:lec_analysis}
In this section, we analyze the stability and performance of the Buck converter featuring the LEC DR scheme defined by Eq.~\eqref{eq:def_estim_lec} and \eqref{eq:def_compens}. We show that introducing LEC does not affect the command-to-output plant transfer function, thus preserving nominal stability. Next, we derive a sufficient condition for robust stability. 

\subsection{Nominal stability and performance}
\label{sec:lec_nominal_analysis}
Consider the inner closed-loop system defined by the plant and the LEC:
\begin{subequations}\label{cl}    
\begin{align}
    \begin{bmatrix} v_o \\ i_L \end{bmatrix} &= P \begin{bmatrix} v_{\rm SW} \\ \tilde i_\textrm{out} \end{bmatrix}, \qquad v_{\rm SW} = \kff (v_c+v_\textrm{inj}) \label{cl_12} \\
    v_\textrm{inj} &= \frac{-G_2}{\kff (1+s/p_H)} ( -G_1 v_o + i_L ). \label{cl_3}
\end{align}
\end{subequations}
The following result holds.

\begin{theorem}[Input-output behavior of the inner loop]
    The system \eqref{cl} is nominally stable, and its input-output behavior is described by
    \begin{equation}\label{nominal_closedloop}
        v_o = P_{11}\kff v_{c} + P_{12} \frac{s}{s+p_H} \tilde i_\textrm{out}.
    \end{equation}
\end{theorem}
\begin{proof}
    From \eqref{cl}, we get
    \begin{align}
        v_o &= P_{11}\kff v_{c} +  P_{11} \frac{G_2}{1+s/p_H} \left( G_1 v_o - i_L \right) + P_{12} \tilde i_\textrm{out} \label{inloop_1} \\
        i_L &= P_{21}\kff v_{c} +  P_{21} \frac{G_2}{1+s/p_H} \left( G_1 v_o - i_L \right) + P_{22} \tilde i_\textrm{out} \label{inloop_2}
    \end{align}
    Next, we get $i_L$ from \eqref{inloop_2} as
    \begin{align}\label{eq:il_th1}
        i_L &= \frac{P_{21}^2 P_{12}}{P_{11}\left(P_{11}(1+s/p_H) + P_{21}P_{12}\right)} v_o +  \\
        &+ \frac{P_{11}(1+s/p_H)}{P_{11}(1+s/p_H) + P_{21}P_{12}} \left( P_{21}\kff v_{c} + P_{22} \tilde i_\textrm{out} \right). \notag 
    \end{align}
    Replacing \eqref{eq:il_th1} into \eqref{inloop_1} and performing suitable simplications yields \eqref{nominal_closedloop}. The feedback loop is internally stable because \eqref{nominal_closedloop} is BIBO stable, and no unstable cancellations occur when forming the loop function.
\end{proof}
    
Eq.~\eqref{nominal_closedloop} allows us to draw additional considerations on the LEC properties. First, we notice that the transfer function between $v_{c}(t)$ and $v_o(t)$ is preserved. Thus, any controller designed to achieve specific input-output performances on the original plant still performs the same when adopting the LEC. Importantly, this includes the nominal stability of the system, which is preserved.
Secondly, the transfer function between $\tilde i_\textrm{out}(t)$ and $v_o(t)$ gets multiplied by $s/(s+p_H)$, which is a high-pass filter with gain $0$ at low frequency and $1$ at high frequency. Thus, independently of $\mathcal{K}$, the current disturbance will be rejected at steady state. Moreover, by letting $p_H \rightarrow \infty$, the effect of the output current is exactly canceled.

To summarize, the LEC is an additional stage that reduces the Buck output impedance (especially at low frequencies) while preserving the control-to-output behavior. 

\subsection{Robust stability}
\label{sec:robustness}
As presented in Sec.~\ref{sec:Hinf_design}, we design the controller $\mathcal{K}$ to account for {robust stability} by verifying this property through $\mu$-analysis. In the remainder of this section, we study under what conditions the introduction of the LEC \eqref{eq:def_estim_lec}, \eqref{eq:def_compens} preserves the robust stability property. 

In the following, we denote by $\hat \cdot$ the quantities calculated using nominal component values, while the absence of the $\hat \cdot$ symbol indicates the value of the actual components. According to the this notation, the LEC is 
\begin{equation}
    v_{\rm inj} = \frac{-\hat G_2}{\hat k_\textrm{FF} (1+s/p_H)} ( -\hat G_1 v_o +  i_L). \label{cl_3_unc}
\end{equation}
To study the error due to the mismatch between $G_1$ and $\hat G_1$, we introduce the following additive uncertainty description
\begin{equation}\label{def:lamb1}
    G_1(s) = \hat G_1(s) + \Lambda(s) \Delta_1(s),
\end{equation}
where $\Lambda(s)$ is a suitable frequency weighting function and $\Delta_1(s)$ is such that $\norm{\Delta_1}_{\Hinf}<1$. 

\begin{remark}\label{rem:bode_envelope_Lambda}
    Computing a tight description of $\abs{\Lambda(i\omega)}$ given bounds on each component value is possible. Specifically, for each fixed $\omega$, $\abs{\Lambda(i\omega)}$ is a polynomial function of the Buck's components. Therefore, computing the Bode envelope of $\Lambda$ can be recast to polynomial optimization problems, which may be solved for a tight upper bound using semidefinite relaxation; see \cite{cerone2020bode} for a detailed discussion.
\end{remark}

Using Eqs.~\eqref{cl_12} and \eqref{cl_3_unc}, we obtain:
\begin{equation}
    v_o = P_{11}\kff v_{c} +  P_{11} \frac{\hat G_2}{1+s/p_H} \frac{\kff}{\hat k_{\rm FF} } \left( \hat G_1 v_o - i_L \right) + P_{12} \tilde i_\textrm{out}. \label{inloop_1_und}
\end{equation}
Moreover, from \eqref{io_fcn_out}-\eqref{deltaP}, we have $i_L = G_1 v_o + \tilde i_\textrm{out}$. Using this to replace $i_L$ into \eqref{inloop_1_und}, we get
\begin{equation}\label{eq:vc1_to_vo_unc_0}\begin{aligned}
    v_o &= P_{11}\kff v_{c} + \\
    &+P_{11} \frac{\hat G_2}{1+s/p_H} \frac{\kff}{\hat k_{\rm FF} } \left( \hat G_1 v_o - G_1 v_o - \tilde i_\textrm{out} \right) + P_{12} \tilde i_\textrm{out}.
\end{aligned} \end{equation}
To study stability, we focus our attention on the input-output relationship $v_{c}(t)$ to $v_o(t)$. For this reason, we neglect the presence of the disturbance $\tilde i_\textrm{out}(t)$. From \eqref{eq:vc1_to_vo_unc_0} and \eqref{def:lamb1}, we have
\begin{equation}
    v_o = P_{11}\kff v_{c} + P_{11} \frac{\hat G_2}{1+s/p_H} \frac{\kff}{\hat k_{\rm FF} } \Lambda \Delta_1 v_o
\end{equation}
which yields:
\begin{equation}\label{eq:vc1_to_vo_unc}
    v_o = \frac{P_{11}\kff}{1-P_{11}\kff W_r \Delta_1} v_{c},\quad
    W_r \doteq \frac{\hat G_2}{\hat k_\textrm{FF} (1+s/p_H)} \Lambda.
\end{equation}

We highlight that $W_r$ depends only on nominal parameter values and $\Lambda$, which can be computed as discussed in Remark~\ref{rem:bode_envelope_Lambda}. Therefore, $\abs{W_r(i\omega)}$ is easily computed using available information. Conversely, $P_{11}$ and $\kff$ in \eqref{eq:vc1_to_vo_unc} are uncertain. Eq.~\eqref{eq:vc1_to_vo_unc} is the description of a plant with nominal value $P_{11}\kff$ and affected by {inverse multiplicative uncertainty} $W_r$. See, e.g., \cite[Section 8.2]{skogestad2005multivariable} for details on this uncertainty description. This observation leads us to the following result.
\begin{theorem}\label{th:rob_stab}
    Let $\mathcal{K}$ be a controller robustly stabilizing the uncertain plant $P_{11} \kff$. Let $S = (1+P_{11}\kff \mathcal{K} G_f)^{-1}$ denote the sensitivity function of the corresponding feedback control system. Robust stability is preserved after the introduction of the LEC defined according to Eq.~\eqref{eq:def_estim_lec} and \eqref{eq:def_compens} if
    \begin{equation}\label{unstr_stab}
        \lVert S P_{11}\kff W_r \rVert_{\mathcal{H}_\infty } < 1.
    \end{equation}
\end{theorem}

\begin{proof}
    The result follows from the application of the small-gain theorem on the $M-\Delta$ structure defined by the inverse additive uncertainty description \eqref{eq:vc1_to_vo_unc}. See \cite[Section 8.6]{skogestad2005multivariable}.
\end{proof}

\begin{remark}
    The condition provided by Theorem \ref{th:rob_stab} is only sufficient for robust stability and may be conservative.  
    We could use $\mu$-analisys \cite{zhou1998essentials} to get a tight result on robust stability considering the entire control architecture. Nonetheless, Theorem \ref{th:rob_stab} is useful during the design phase to get additional information on the choice of $p_H$ because it is involved in the definition of $W_r$ in Eq.~\eqref{eq:vc1_to_vo_unc}. 
\end{remark}

Condition \eqref{unstr_stab} involves the presence of $P_{11}, \kff$, and $S$, which depend on the \textit{actual} parameters, not the nominal ones. Consequently, we cannot check this condition directly. To proceed with our analysis, we introduce the function $N(\omega)$ with the property $N(\omega) \leq \kff^{-1}\lvert P_{11}(i\omega) S(i\omega) \rvert^{-1}$ for all $\omega \geq 0$.
Similarly to what we discussed in Remark~\ref{rem:bode_envelope_Lambda}, the procedure presented in \cite{cerone2020bode} applies to the problem of computing the function $N(\omega)$ given bounds on the component values.
Using $N(\omega)$, a sufficient condition for \eqref{unstr_stab} is 
\begin{equation}\label{eq:rob_stab_cond3}
    \lvert W_{r}(i\omega) \rvert < N(\omega), \quad \forall \omega.
\end{equation}
Replacing $W_r$, we obtain an explicit condition on the choice of the frequency $p_H$ as
\begin{equation}\label{eq:rob_stab_cond4}
    \abs{ \frac{1}{1+\frac{i\omega}{p_H}} }  < N(\omega) \abs{\frac{\hat k_{\rm FF} }{\hat G_2(i\omega)}} \abs{\frac{1}{\Lambda(i\omega)}}. 
\end{equation}
The left-hand side of \eqref{eq:rob_stab_cond4} represents the magnitude frequency response of a first-order low-pass filter, which has a pole located at $p_H$ and unitary DC-gain. In contrast, the right-hand side is influenced by the components' uncertainties. When $p_H$ is at low frequencies, the left-hand side decreases, making it easier to satisfy \eqref{eq:rob_stab_cond4}. Conversely, if $p_H$ approaches infinity, robust stability is guaranteed if the right-hand side does not exceed one at all frequencies. %However, the latter situation is often unrealistic because high-frequency effects are typically neglected, meaning that the uncertainty will be significant at higher frequencies.

To summarize, the pole at $p_H$ renders the system physically realizable and enhances the system's robust stability at the cost of a performance decrease. Eq.~\eqref{eq:rob_stab_cond4} offers a quantitative way of tuning $p_H$ to trade off nominal performances and stability robustness. We conclude this section with another useful observation on the nature of the uncertainty. Since $G_1$ only depends on components $R_L, R_C, C$, a huge uncertainty on their value potentially reduces the stability robustness, while uncertainty on $L, R_i, R_{\rm ON}^{\rm HS, LS}$ (which defines $G_2$) affects the stability robustness only \textit{indirectly} through the factor $N(\omega)$. In other words, if $R_L, R_C, C$ are precisely known, then $\abs{\Lambda}(i\omega) \approx 0$ and condition \eqref{eq:rob_stab_cond4} is fulfilled for any value of $L,R_i,R_{\rm ON}^{\rm HS, LS}$. This indicates that, during Buck design, the robust stability benefits from choosing a more expensive capacitor, while the precision of the inductor is almost irrelevant.

%%%%%%%%%%%%%%%%%%%%%%%%%%%%%%%%%%%%%%%%%%%%%%%%%%%%%%%%%%%%%%%%%%%%%%%

\section{Experimental results}
\label{sec:experim_res}

In this section, we compare the performance of LEC, DOB, and UIO through numerical simulations. Then, we discuss the analog circuit implementation of LEC and perform experimental measurements to validate the effectiveness of the proposed LEC-based approach on a real Buck converter. In the following, we will refer to the Buck converter parameter values in Table~\ref{tab:case_study}. 

\begin{figure}
    \centering
        \input{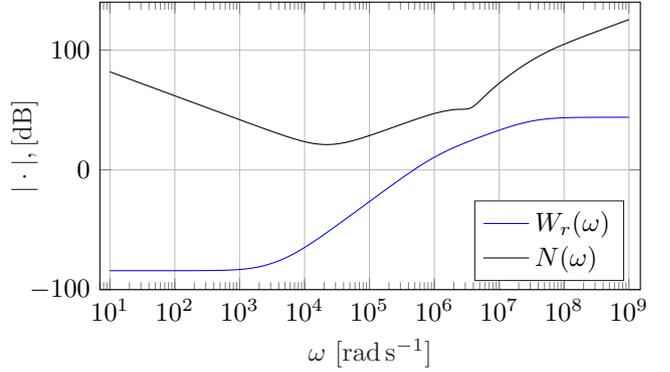}
    \caption{Magnitude Bode diagram of $W_1$ and $N$ to check the robust stability condition \eqref{eq:rob_stab_cond3}}
    \label{fig:robBode}
\end{figure}

\begin{figure*}
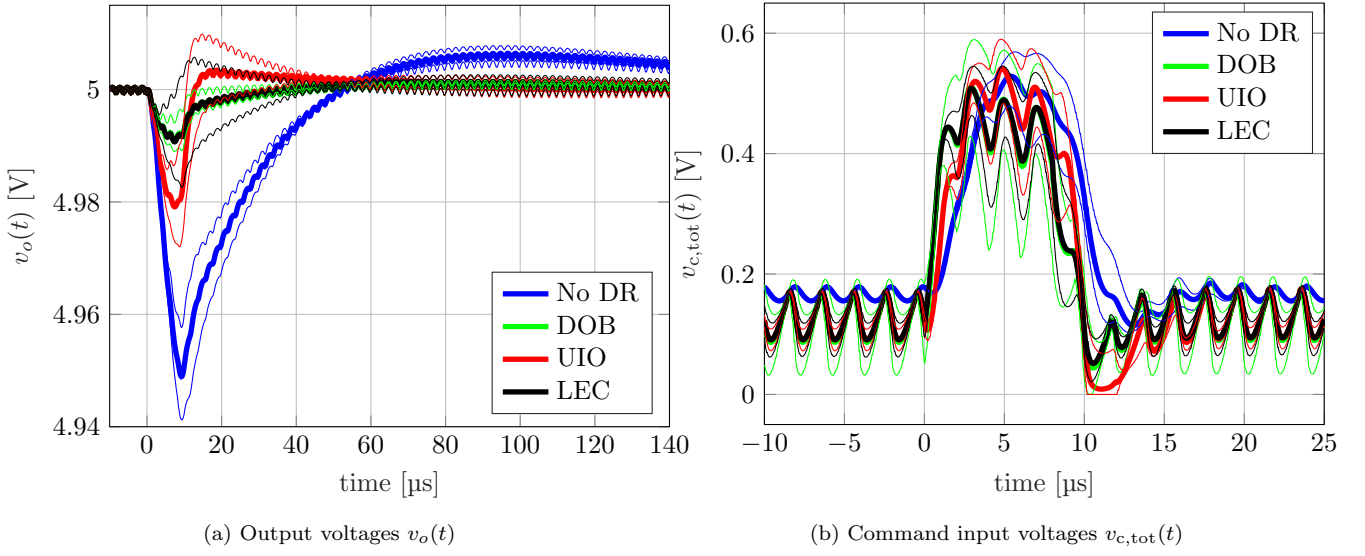

    \begin{subfigure}{0.5\textwidth}
        \centering
        \input{montecarlo_vo_50}
        \caption{Output voltages $v_o(t)$}
        \label{fig:simul50_vo}
    \end{subfigure}
    \hfill 
    \begin{subfigure}{0.5\textwidth}
        \centering
            \input{montecarlo_vc_50}
        \caption{Command input voltages $v_{\rm c,tot}(t)$}
        \label{fig:simul50_vc}
    \end{subfigure}
    \caption{Comparison between uncompensated (blue), LEC (black), DOB (green), and UIO (red). The thick line is the average response among 50 runs, while the thin ones are bounds over all runs.}
    \label{fig:sim_mc}
\end{figure*}

\begin{figure*}
    \centering
        \includegraphics[width = 1.9\columnwidth]{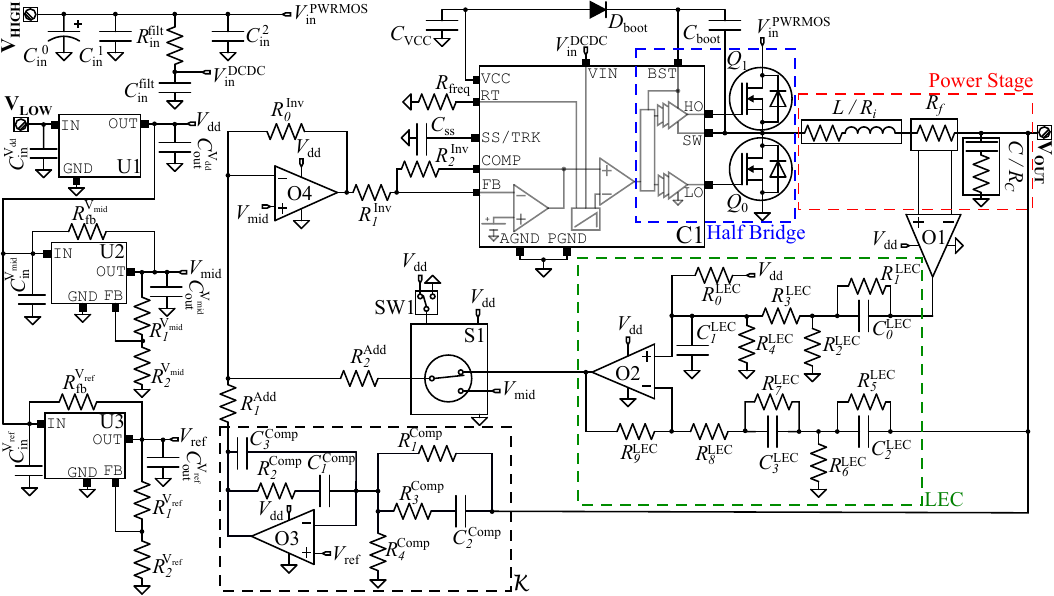}
    \caption{VMC Buck converter embedding the proposed LEC. The LEC is implemented as an analog circuit network.}
    \label{fig:lec_circuit}
\end{figure*}

\begin{figure}
    \centering
    \includegraphics[width=.9\columnwidth]{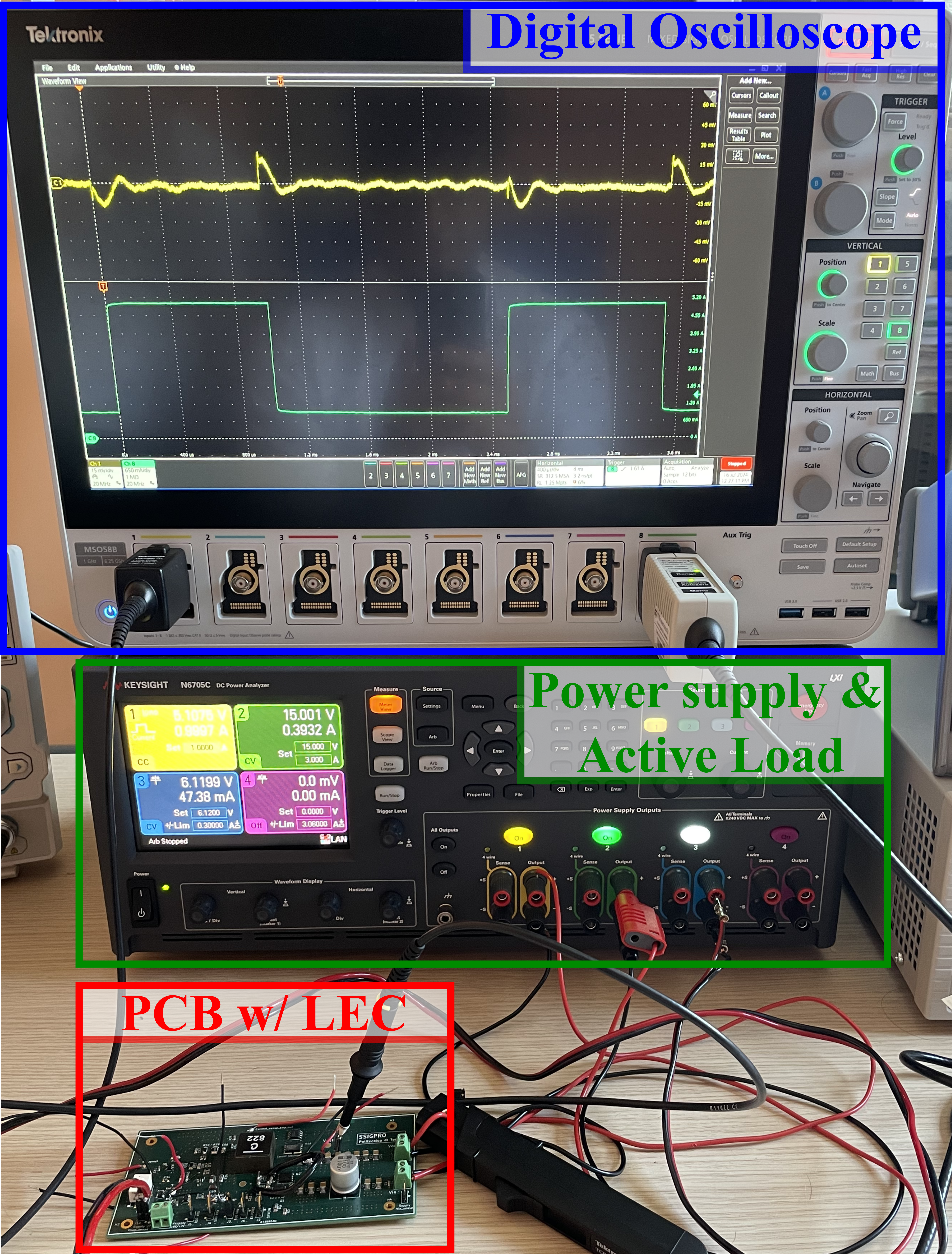}
    \caption{Experimental measurement setup.}
    \label{fig:lab_setup}
\end{figure}

\begin{figure*}[h]
    \centering
    \includegraphics[width=0.8\linewidth]{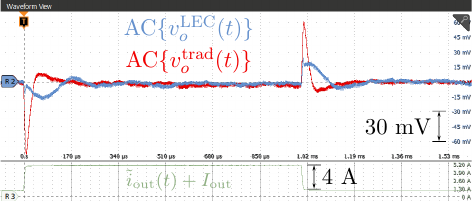}
    \caption{Oscilloscope measurements from prototype PCB.}
    \label{fig:oscilloscope}
\end{figure*}

\begin{table*}
    \centering
    \caption{Values and part number of the circuit elements  in Fig.~\ref{fig:lec_circuit}.}
    \label{tab:component_values}
    \resizebox{\textwidth}{!}{%
    \begin{tabular}{c c |@{\quad}| c c |@{\quad}| c c}
        \hline
        {$\begin{array}{c}
             \textbf{Schematic}  \\
             \textbf{Label} 
        \end{array}$} & {$\begin{array}{c}
             \textbf{Part Number /}  \\
             \textbf{Component Value} 
        \end{array}$} & 
        {$\begin{array}{c}
             \textbf{Schematic}  \\
             \textbf{Label} 
        \end{array}$} & {$\begin{array}{c}
             \textbf{Part Number /}  \\
             \textbf{Component Value} 
        \end{array}$} & 
        {$\begin{array}{c}
             \textbf{Schematic}  \\
             \textbf{Label} 
        \end{array}$} & {$\begin{array}{c}
             \textbf{Part Number /}  \\
             \textbf{Component Value} 
        \end{array}$} \\ 
        \hline

        $C_{\rm in}^{0}$ & \SI{100}{\micro \farad} & 
        $C_{\rm in}^{\rm 1}$ & $2$ x $\SI{10}{\micro \farad}$ & $C_{\rm in}^{\rm 2}$ & $10$ x $\SI{10}{\micro \farad}$  \\ 
        $R_{\rm in}^{\rm filt}$ & \SI{2.2}{\ohm}&
        $C_{\rm in}^{\rm filt}$ & \SI{.1}{\micro \farad} \\ [0.5ex]
        
        \hline

        $C_{\rm in}^{V_{\rm dd}}$ & \SI{1}{\micro \farad} 
        & $C_{\rm out}^{V_{\rm dd}}$ & \SI{4.7}{\micro \farad} 
        & $C_{\rm in}^{V_{\rm mid}}$ & \SI{1}{\micro \farad} \\ [0.5ex] 
        $C_{\rm out}^{V_{\rm mid}}$ & \SI{4.7}{\micro \farad} 
        & $C_{\rm in}^{V_{\rm ref}}$ & \SI{1}{\micro \farad} & $C_{\rm out}^{V_{\rm ref}}$ & \SI{4.7}{\micro \farad} \\ [0.5ex]        
        $R_{\rm fb}^{V_{\rm ref}}$ & \SI{200}{\micro \ohm} 
        & $R_{2}^{V_{\rm ref}}$ & \SI{10}{\kilo \ohm} 
        & $R_{1}^{V_{\rm ref}}$ & \SI{3.4}{\kilo \ohm} \\ [0.5ex]
        $R_{\rm fb}^{V_{\rm mid}}$ & \SI{200}{\ohm} 
        & $R_{2}^{V_{\rm mid}}$ & \SI{12}{\kilo \farad} 
        & $R_{1}^{V_{\rm mid}}$ & \SI{38}{\kilo \farad} \\ [0.5ex]
        U1 & LDL1117S50R 
        & U2 & CAT102TDI-GT3 
        & U3 & CAT102TDI-GT3 \\ [0.5ex]

        \hline

        $L/R_i$ & XAL1510-822MED / \SI{8.2}{\micro\henry} & 
        $R_f$ & WSHM2818R0150FEA / \SI{15}{\milli \ohm} & $C/R_{C_o}$ & GRM32EC72A106KE05 / $26$ x $\SI{10}{\micro \farad}$  \\ [0.5ex] 

        \hline

        $Q_1$ &  STL90N10F7 & $Q_0$ &  STL110N10F7 & & \\[0.5ex] 
        
        \hline %%
        
        C1 &  L3751PUR & $R_{\rm freq}$ &  \SI{20}{\kilo \ohm} & $C_{\rm ss}$ & \SI{47}{\nano \farad} \\ [0.5ex] 
        $C_{\rm VCC}$ & \SI{1}{\micro \farad}  & $D_{\rm boot}$ & BAT46ZFILM & $C_{\rm boot}$ & \SI{100}{\nano \farad}\\ [0.5ex]
        $R_1^{\rm Inv}$ &  \SI{10}{\kilo \ohm} & $R_2^{\rm Inv}$ &  \SI{10}{\kilo \ohm} &&  \\ [0.5ex]  
         
        \hline

        O1 &  AD8410AWBRZ-RL & O2 & ADA4891 & $R_0^{\rm LEC}$ & \SI{68}{\kilo \ohm} \\ [0.5ex]
        $R_1^{\rm LEC}$ & \SI{1}{\mega \ohm} & $R_2^{\rm LEC}$ & \SI{4.12}{\kilo \ohm} & $R_3^{\rm LEC}$ & \SI{68}{\kilo \ohm} \\ [0.5ex]
        $R_4^{\rm LEC}$ & \SI{680}{\kilo \ohm} & $R_5^{\rm LEC}$ & \SI{1}{\mega \ohm} & $R_6^{\rm LEC}$ & \SI{35}{\kilo \ohm} \\ [0.5ex]
        $R_7^{\rm LEC}$ & \SI{10}{\mega \ohm} & $R_8^{\rm LEC}$ & \SI{316}{\ohm} & $R_9^{\rm LEC}$ & \SI{350}{\kilo \ohm} \\ [0.5ex]
        $C_0^{\rm LEC}$ & \SI{470}{\pico \farad} & $C_1^{\rm LEC}$ & \SI{1}{\nano \farad} & $C_2^{\rm LEC}$ & \SI{470}{\pico \farad} \\ [0.5ex]
        $C_3^{\rm LEC}$ & \SI{100}{\pico \farad} && \\ [0.5ex]

        \hline

        O3 & ADA4891 & $R_1^{\rm Comp}$ & \SI{1}{\mega \ohm} 
        & $R_2^{\rm Comp}$ & \SI{4.12}{\kilo \ohm} \\ [0.5ex]
        $R_3^{\rm Comp}$ & \SI{68}{\kilo \ohm} & $R_4^{\rm Comp}$ & \SI{680}{\kilo \ohm} & $C_1^{\rm Comp}$ & \SI{5.6}{\nano \farad} \\ [0.5ex]
        $C_2^{\rm Comp}$ & \SI{1.8}{\nano \farad} & $C_3^{\rm Comp}$ & \SI{150}{\pico \farad} \\ [0.5ex]

        \hline

        O4 & ADA4891 & S1 & ADG819BRZ & SW1 & 204-121 \\ [0.5ex]
        $R_0^{\rm Add}$ & \SI{10}{\kilo \ohm} & $R_1^{\rm Add}$ & \SI{10}{\kilo \ohm} & $R_2^{\rm Add}$ & \SI{10}{\kilo \ohm} \\ [0.5ex]

        \hline
        
    \end{tabular}
    }
\end{table*}

\subsection{Simulation results}
\label{sec:simul}

The LEC is designed as described in Sec.~\ref{sec:lec} and \ref{sec:comp_design}. The only free parameter to be selected, $p_H$, is set to $p_H=10^6\,\textrm{rad}\,\textrm{s}^{-1}$ to trade-off DR performance and robustness. In Fig.~\ref{fig:robBode}, we show the Bode diagram of $W_1(s)$ and $N(s)$, from which we conclude that condition~\eqref{eq:rob_stab_cond4} is fulfilled and, thus, robust stability is guaranteed. For a fair comparison, we set $p_H=10^6\,\textrm{rad}\,\textrm{s}^{-1}$ in~\eqref{eq:Gdob} for the DOB design and $\lambda_{1}=\lambda_2=10^6\,\textrm{rad}\,\textrm{s}^{-1}$ and $\lambda_3 = 0.95\lambda_1$ in~\eqref{eq:lamb_uio} for the UIO design.

We perform $50$ simulations, selecting different values for each component. We set $V_{\rm in} = \SI{20}{\volt}$ and the $R_L=\SI{5}{\ohm}$ to achieve a static $\SI{1}{\ampere}$ load current value. All other component values are sampled randomly from the uniform distribution in the respective uncertainty interval. In each simulation, we apply a current disturbance step with an amplitude of $\SI{8}{\ampere}$ and a slope of $\SI{1e6}{\ampere\per\second}$. 

Fig.~\ref{fig:simul50_vo} and~\ref{fig:simul50_vc} show the responses of the output voltage $v_o(t)$ and the command input signal $v_{\rm c, tot}(t)$, respectively. Specifically, we compare the performance without the DR stage against the ones achieved using LEC, DOB, and UIO. Solid lines represent the average of all the simulation outcomes, while thin lines represent upper and lower bounds over all the realizations.

All the examined techniques improve the DR performances compared to uncompensated VMC. Among the three analyzed schemes, UIO shows the worst performance concerning the response to the disturbance and the command activity, which saturates for several switching periods in several simulations. On average, DOB and LEC enjoy the same load DR performance regarding overshoot and settling time. Compared to LEC, the DOB scheme enjoys a lower variance in the output response. However, it also features a larger variance of the command input $v_\textrm{c,tot}(t)$, due to the overcompensation effect. Indeed, in some simulations, the DOB command activity saturates, while the LEC one does not.

\subsection{Circuital implementation}
\label{sec:circuital_implentation}
To validate the effectiveness of the proposed LEC, we implemented a custom analog circuit through a Printed Circuit Board (PCB). The system implements a VMC Buck converter with LEC DR, as schematized in Fig.~\ref{fig:circuit} from a high-level perspective. The circuit-level implementation of the complete system is shown in Fig.~\ref{fig:lec_circuit}. The detailed list of the part numbers and the values of the components employed in the schematic is available in Table~\ref{tab:component_values}. In the design phase, we referred to the main Buck converter parameters listed in Table~\ref{tab:case_study}. 

The HB is realized via the power MOSFETs $Q_{0,1}$, while the PS includes both the converter output filter, consisting of the $L/R_i$ and $C/R_{C}$ elements, and the inductor current sensing resistance $R_f$. The inductor current signal is conditioned via both $R_f$ and the current-sense amplifier O1.
The input voltage of the converter is supplied through the external pin $\rm V_{HIGH}$, which is connected to a network of decoupling capacitors constituted by $C_{\rm in}^{0,1,2}$. The converter output is available on the external pin $\rm V_{OUT}$. The MOSFETs $Q_{0,1}$ are driven by the synchronous Buck controller C1, which is supplied from the external input voltage on $\rm V_{HIGH}$ additionally filtered by $R_{\rm in}^{\rm filt}-C_{\rm in}^{\rm filt}$. 

The controller embeds the gate drivers, the sawtooth-based PWM stage, and additional circuit blocks implementing the VMC. The switching frequency is set via the resistor $R_{\rm freq}$, while the soft-start time is programmable via the capacitor $C_{\rm ss}$. An internal linear regulator generates a reference voltage on the \texttt{VCC} pin, which serves to implement the Dickson charge pump $D_{\rm boot}$-$C_{\rm boot}$. The outer loop controller $\mathcal{K}$ is implemented through a type-III compensation network highlighted in black in Fig.~\ref{fig:lec_circuit}. The LEC circuit implementation design is conducted by considering the cascaded connection of the $\mathcal{E}$ and $\mathcal{F}$ blocks. It consists of one OpAMP and passive components and is highlighted in green in Fig.~\ref{fig:lec_circuit}. 

To either enable or disable the operation of the LEC, the circuit is equipped with a solid-state switch S1, which is in turn driven from a mechanical switch SW1. When the LEC is disabled, the feedback control action is uniquely provided by $\mathcal{K}$, implementing the traditional VMC technique. Otherwise, the LEC output is added to the $\mathcal{K}$ output, thus realizing the proposed load DR mechanism. The output signals of the LEC and $\mathcal{K}$ are added via the analog stage that comprises $R_1^{\rm Add}$, $R_2^{\rm Add}$, $R_0^{\rm Inv}$ and O4. Resistors $R_1^{\rm Inv}$ and $R_2^{\rm Inv}$ are required to configure the error amplifier inside C1 as an inverting stage with unitary gain. This inverting stage is required because the internal error amplifier is physically connected to the PWM stage. The stages U1, U2 and U3 together with their feedback resistors (i.e., $R_{1,2}^{V_{\rm mid,ref}}$ and $R_{\rm fb}^{V_{\rm mid,ref}}$) and decoupling capacitors (i.e., $C_{\rm in,out}^{V_{\rm dd}}$) are voltage regulators and references. These are required to generate the voltage levels that are employed to both supply the active analog stages in the system (i.e., the OpAmps and the current-sense amplifier) and set the voltage reference values (e.g., the reference on the + terminal of O3 for $\mathcal{K}$). These stages are supplied via the external voltage applied on the pin $\rm V_{LOW}$, which is in the range $[6, 7]\SI{}{\volt}$. 

\subsection{Experimental measurements}
\label{sec:exp_measurements}
Experimental measurements on the PCB prototype described in Sec.~\ref{sec:circuital_implentation} have been conducted on the circuit shown in Fig.~\ref{fig:lec_circuit}. These validate the effectiveness of the LEC in mitigating external load disturbances. 

The experimental setup is shown in Fig.~\ref{fig:lab_setup}. The Buck converter is supplied via the DC power module Keysight N6752A, which is installed in the Keysight N6705C DC power analyzer. We apply a fixed resistive load $R_L$ which fixes the static converter output current value to $\SI{1}{\ampere}$. We examined the load transient response of the converter when a sequence of step-load changes with amplitude $\SI{4}{\ampere}$ is applied. The step-load is provided via the Keysight N6791A DC electronic load module installed in the Keysight N6705C. We use the Tektronix MSO58B oscilloscope to measure the load current and output voltage, both considering operation with disabled and enabled LEC. 
The comparative results are shown in Fig.~\ref{fig:oscilloscope}, showing the AC-coupled $v_o(t)$ and the applied disturbance $\tilde{i}_{\rm out}(t)$. When the proposed LEC compensation is active, the output voltage undershoot/overshoot is significantly reduced. Specifically, the undershoot is $\hat{s}_\textrm{trad} = \sfrac{e_\textrm{max}^{\rm trad}}{V_o^{\textrm{target}}} = \sfrac{\SI{70}{\milli\volt}}{\SI{5}{\volt}} = 1.4\%$ and $\hat{s}_\textrm{LEC} = \sfrac{e_\textrm{max}^{\rm LEC}}{V_o^{\textrm{target}}} = \sfrac{\SI{18}{\milli\volt}}{\SI{5}{\volt}} = 0.36\%$ when LEC is respectively disabled and enabled.

\section{Conclusion}
\label{sec:concl}
This study presents three novel disturbance rejection mechanisms for load current disturbances in DC-DC Buck converters. Two of the proposed techniques build on well-established methods: the Disturbance Observer (DOB) and the Unknown Input Observer (UIO). The third one, named Load Estimator-Compensator (LEC), is an original approach tailored for Buck converters. These methods enhance disturbance rejection by augmenting the controller action and are designed for seamless integration into conventional control loops.

We conduct a rigorous stability analysis that ensures robust stability of the overall system when integrating the LEC. Simulations validate the effectiveness of the proposed approaches, showing that DOB and LEC achieve similar average performances. While DOB is less sensitive to parameter uncertainty, it requires more aggressive control, potentially leading to saturation effects. This motivates further investigations on this approach.

Experimental validation on a custom PCB implementing a voltage-mode Buck converter with LEC mechanism confirms the effectiveness of the proposed solution and corroborates its practical viability for high-performance applications.

\bibliographystyle{ieeetr}
\bibliography{references}

\end{document}